\documentclass[preprint]{aastex}
\usepackage{amssymb, amsmath, amsthm}
\RequirePackage{xspace}
\shorttitle{Post-photospheric resonance-scattering in SNe}
\shortauthors{Friesen et al.}
\def\ifundefined#1{\expandafter\ifx\csname#1\endcsname\relax}

\ifundefined{ensuremath}\def\ensuremath#1{\relax\ifmmode{#1}}
\else${#1}$\fi\else\relax\fi
\ifundefined{nuc}\def\nuc#1#2{\relax\ifmmode{}^{#1}{\protect\mathrm{#2}}
\else${}^{#1}$#2\fi}\else\relax\fi

\newcommand{\bea}{\begin{eqnarray}}
\newcommand{\eea}{\end{eqnarray}}

\newcommand{\kmps}{\ensuremath{\mathrm{km}~\mathrm{s}^{-1}}\xspace}

\newcommand{\nco}{\ensuremath{\nuc{56}{Co}}\xspace}

\newcommand{\synow}{\texttt{SYNOW}\xspace}

\newcommand{\snia}{SN~I\lowercase{a}\xspace}
\newcommand{\sneia}{SNe~I\lowercase{a}\xspace}

\newcommand{\zres}{\ensuremath{z_{\mathrm{res}}}}
\newcommand{\mures}{\ensuremath{\mu_{\mathrm{res}}}}
\newcommand{\rc}{\ensuremath{r_{c}}}
\newcommand{\rmax}{\ensuremath{r_{s}}}

\graphicspath{{./}}
\DeclareGraphicsExtensions{.eps,.EPS}

\begin{document}

\title{Supernova Resonance--scattering Line Profiles in the Absence of a
Photosphere}

\author{Brian Friesen\altaffilmark{1}, E.~Baron\altaffilmark{1,2,3,4},
  David Branch\altaffilmark{1}, Bin Chen\altaffilmark{1},
  Jerod~T.~Parrent\altaffilmark{5,6}, and
  R.~C.~Thomas\altaffilmark{3}}

\altaffiltext{1}{Homer L.~Dodge Department of Physics \& Astronomy,
  University of Oklahoma, 440 W. Brooks St., Rm. 100, Norman, OK
  73019, USA}

\altaffiltext{2}{Hamburger Sternwarte, Gojenbergsweg 112, 21029
  Hamburg, Germany}

\altaffiltext{3}{Computational Cosmology Center, Computational
  Research Division, Lawrence Berkeley National Laboratory, MS
  50B-4206, 1 Cyclotron Road, CA 94720, USA}

\altaffiltext{4}{Department of Physics, University of California,
  Berkeley, CA 94720, USA}

\altaffiltext{5}{6127 Wilder Lab, Department of Physics \& Astronomy,
  Dartmouth College, Hanover, NH 03755, USA}

\altaffiltext{6}{Las Cumbres Observatory Global Telescope Network,
  Goleta, CA 93117, USA}

\begin{abstract}
In supernova spectroscopy relatively little attention has been given
to the properties of optically thick spectral lines in epochs
following the photosphere's recession.  Most treatments and analyses
of post-photospheric optical spectra of supernovae assume that
forbidden-line emission comprises most if not all spectral
features. However, evidence exists which suggests that some spectra
exhibit line profiles formed via optically thick resonance-scattering
even months or years after the supernova explosion.  To explore this
possibility we present a geometrical approach to supernova spectrum
formation based on the ``Elementary Supernova'' model, wherein we
investigate the characteristics of resonance-scattering in optically
thick lines while replacing the photosphere with a transparent central
core emitting non-blackbody continuum radiation, akin to the optical
continuum provided by decaying \nco formed during the explosion.  We
develop the mathematical framework necessary for solving the radiative
transfer equation under these conditions, and calculate spectra for
both isolated and blended lines. Our comparisons with analogous
results from the Elementary Supernova code \synow reveal several
marked differences in line formation. Most notably, resonance lines in
these conditions form P~Cygni-like profiles, but the emission peaks
and absorption troughs shift redward and blueward, respectively, from
the line's rest wavelength by a significant amount, despite the
spherically symmetric distribution of the line optical depth in the
ejecta.  These properties and others that we find in this work could
lead to misidentification of lines or misattribution of properties of
line-forming material at post-photospheric times in supernova optical
spectra.
\end{abstract}

\section{Introduction}
Many of the physical processes which contribute to spectrum formation
in a supernova (SN) change dramatically as it ages. Given the
difficulty and complexity of including all such processes, as well as
their evolution in time, exploration into this computational frontier
has begun only very recently \citep{PE00, h03a, kasen06a, jhb09,
  kromersim09, soma10a, DH10a, dh11, HD12}.  Most work in SN
spectroscopy has focused on the early ``photospheric'' phase,
comprising the time from explosion to a few weeks post-maximum light,
where resonance-scattering from permitted lines dominates the spectrum
\citep{bran81b, branch81b85, jeffetal92, ML93, mazzali95, fish90n97,
  mazz91bg97, hatano94D99, mazzcode00, mazz90N01, branchcomp105,
  dessart05a, mazzali05a, hach06}.  Considerably less attention has
been given to the ``nebular'' phase, several months or more after
explosion, where emission from forbidden lines constitute most of the
spectrum \citep{axelphd80, rplucy92, kozfran98a, kozfran98b,
  mazzali98, maeda98bw06, jerkstrand11, mazz03hv11}.  Finally, the
intermediate phases, that is, a few months post-explosion, have
received the least amount of scrutiny \citep{Maurer11}.  One reason so
much attention focuses on either very early or very late times is that
in these regimes one can reproduce with reasonable fidelity the
physical processes which dominate spectrum formation through a number
of simplifying assumptions.  For example, in the early, photospheric
phase the Sobolev approximation \citep{castor70} and a
resonance-scattering source function \citep{jb90} are both accurate
approximations due to the high densities and steep velocity gradients
in the SN ejecta; at late times one typically assumes both that the
ejecta is optically thin and that line emission arises exclusively
from forbidden lines \citep{axelphd80}.  However these two groups of
assumptions are generally incompatible with each other in the
intermediate regime of a SN.

\begin{figure}
\centering
  \includegraphics[scale=0.6]{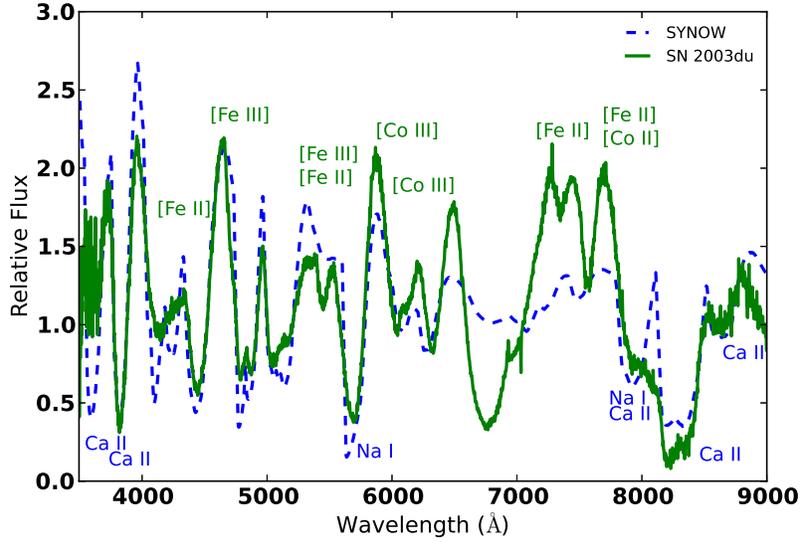}
  \caption{A \synow fit to the +87 day spectrum of SN 2003du. The
    synthetic spectrum contains Na I, Ca II, and Fe II, each with an
    excitation temperature of 7000 K. The photospheric velocity is
    7000 \kmps, and the spectrum has been divided through by the
    blackbody continuum. The permitted line identifications (in blue)
    are from the \synow fit, with Fe II features unlabeled.  The
    alternative forbidden line IDs (in green) follow those in other
    \sneia made by \citet{bowersetal97}.}
\label{fig:synow_03du}
\end{figure}

\begin{figure}
\centering
  \includegraphics[scale=0.6]{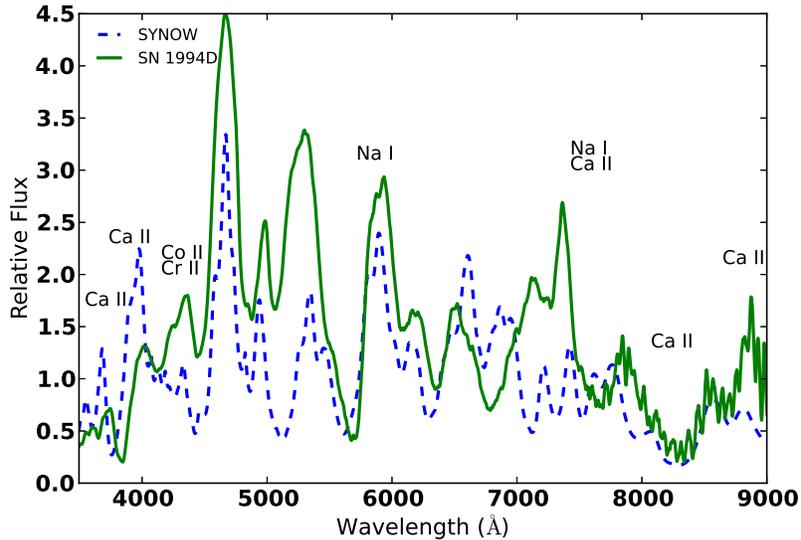}
  \caption{A \synow fit to the +115 day spectrum of SN 1994D. The
    synthetic spectrum contains Na~I, Ca II, Cr II, Fe II, and Co II,
    each with an excitation temperature of 7000 K. The photospheric
    velocity is 6000 \kmps, and the spectrum has been divided through
    by the blackbody continuum. As in Figure~\ref{fig:synow_03du} the
    unlabeled features in the \synow spectrum are due to Fe~II.}
\label{fig:synow_94D}
\end{figure}

Complicating matters further is the possibility that the evolution of
the different types of line-forming processes is asynchronous.
Specifically, the emergence of forbidden emission lines in a SN
spectrum may not herald the systematic withdrawal of
resonance-scattering in optically thick permitted lines. Though this
possibility has frequently been acknowledged \citep{bowersetal97,
  branchcomp105, jerkstrand11}, it has to our knowledge never been
pursued in detail until now.  To illustrate this point we show in
Figures~\ref{fig:synow_03du} and \ref{fig:synow_94D} \synow fits to
the day +87 optical spectrum of SN 2003du and the day +115 spectrum of
SN 1994D, respectively. In the SN 2003du fit we use only three ions:
Na I, Ca II and Fe II. The photospheric velocity is 7000 \kmps, and
the photospheric temperature is 8000 K.  The excitation temperatures
of all three ions is 7000 K. In the SN 1994D fit, we include five
ions: Na I, Ca II, Cr II, Fe II, and Co II. Here the photospheric
velocity is 6000 \kmps and the temperature is 10,000 K. The excitation
temperature of all ions is 7000~K. The observed and synthetic spectra
of both SN 1994D and SN 2003du have been flattened using the method of
\cite{jeffery07a}.

The fits to the observed spectra in both of these figures are
relatively good, and in both fits only permitted lines were
considered.  Even among the features which \synow cannot reproduce
accurately, most of the emission peaks and absorption troughs in the
synthetic spectra form at the same wavelengths as in the observed
spectra, and only the strengths of the features are disparate. (The
exceptions to this are the wavelength range 6600~--~7800 \AA\ in SN
2003du and 6600~--~7500~\AA\ in SN 1994D, throughout which \synow
fails completely to reproduce the observed features.)  It would
therefore be a remarkable coincidence if the observed features in
these two objects arise from purely forbidden emission
\citep{branchcomp105}.

In addition to the circumstantial evidence for persistent
resonance-scattering provided by these reasonably good \synow fits to
several-months-old type Ia supernovae (\sneia), calculations by
\cite{atlas99} and \cite{jerkstrand11} further defend this
claim. \cite{atlas99} calculated the Sobolev optical depth (Equation~1 of
that paper) for the most common ions observed in \snia optical
spectra. In Figure~9c of that work they show the Sobolev LTE optical
depth for Fe I \& II and Co I \& II for an iron-rich composition at 80
days post-explosion in a \snia model, and find that the optical depths
for those ions can be as high as $\tau \sim 50,000$.  Similarly, the
simplified form of the Sobolev optical depth shown in Equation~39 of
\cite{jerkstrand11} shows that the optical depths for some UV
resonance lines (e.g., Mg II $\lambda 2798$) can still be as high as
$\tau \sim 1000$ even eight \emph{years} post-explosion.

In light of the evidence presented above for resonance-scattering at
relatively late times in \sneia, we present a geometrical exploration
of this phenomenon in the spirit of the Elementary Supernova approach
of \citet{jb90}. Rather than attempt to discern exactly how late after
a SN explosion resonance-scattering continues to contribute
significantly to spectrum formation, we assume simply that the
photosphere has receded and that the continuum opacity in the core of
the SN is negligibly small.  We replace the photosphere with a
transparent core emitting non-blackbody continuum radiation and
distribute ions both inside this ``glowing'' core and outside the core
in a non-emitting shell. We then calculate emergent flux profiles for
lines with parameterized optical depths in several different
geometrical configurations. We scrutinize line formation and behavior
in both isolated and blended cases.  Because \synow inspired this
model, the two naturally invite comparison, which we indulge
throughout this work. In \S\ref{sec:core_continuum} we treat the case
of pure continuum (no lines) in the core; in \S\ref{sec:line_in_core}
we introduce a single line in the core; in
\S\ref{sec:one_line_outside_core} we surround the glowing core with a
transparent, non-glowing shell and move the line to the shell; in
\S\ref{sec:two_lines_in_core} we account for the effects of multiple
scattering by placing two lines in the core; \S\ref{sec:disc} contains
our discussion of the implications of our results on interpretation of
late-time spectra of SNe; in \S\ref{sec:conc} we conclude our
work. Finally, we include Appendices which contain complete
mathematical derivations so that all of our work may easily be
reproduced.

\section{A transparent, uniformly emitting core}
\label{sec:core_continuum}

We begin by assuming the SN is spherically symmetric.  Under this
circumstance it is natural to work in $(p,z)$ coordinates, where $p$
is the impact parameter of a ray relative to the center of the SN,
such that the $p = 0$ ray exactly bisects the SN; and $z$ is
orthogonal to $p$, with the $z = 0$ line also exactly bisecting the
SN.  We will at times transform to spherical polar coordinates for
computational expedience, where the radial coordinate $r$ satisfies
\begin{equation}
  \label{eq:r_of_p_and_z}
  r^2=p^2+z^2.
\end{equation}
We also assume that the SN undergoes homologous expansion, $v = r/t$,
so that surfaces of constant line-of-sight velocity are vertical
planes, that is, planes of constant $z$.  We further assume that the
observer is located at $z \rightarrow -\infty$, in which case all rays
incident on the observer are parallel.  Next, we assume that the
photosphere has receded and has been replaced by a transparent,
spherically symmetric core with outer radius $\rc$ which emits
continuum (but not blackbody) radiation. The post-photospheric spectra
of many \sneia contain a flat and weak continuum which is either
thermal in nature or, as \cite{bowersetal97} suggest, due to the sea
of weak optical lines of lowly ionized Fe and Co. At very late times
when the spectrum becomes truly nebular, there is no optical thermal
continuum, but we do not seek to extend our methods into this very
late regime.  To mimic this pseudo-continuum we assign to the core a
grey, spatially constant volume emissivity $j_\lambda(r) = j$ for all
radii $r \leq \rc$ and all wavelengths $\lambda$.  This assumption
allows us to present simple analytic results; with only slight
modification, our methods for calculating line profiles are amenable
to chromatic emissivity.

\subsection{Continuum only}
\begin{figure}
\centering
  \includegraphics[scale=0.45]{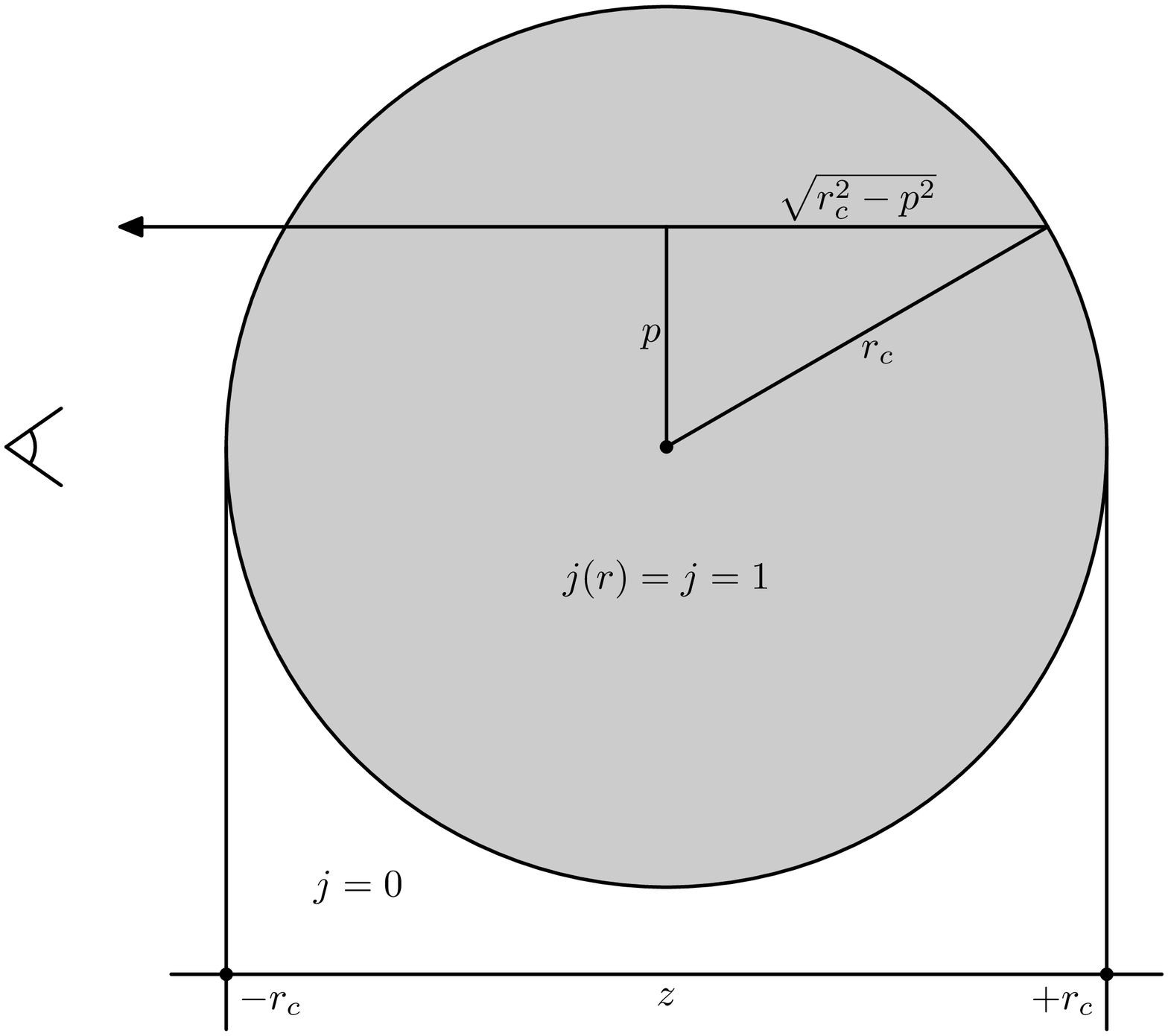}
  \caption{A continuum ray in the core of the SN. In this figure and
    all which follow, we use grey shading to label the
    continuum-emitting region.}
  \label{fig:core_single_ray_with_z_axes}
\end{figure}

In general, to calculate the emergent flux from a SN atmosphere one
must first calculate the source function at all locations in the SN,
followed by the emergent intensity of rays exiting the ejecta toward
the observer.  However, since in this model we neglect all continuum
opacity, the source function is not a well defined quantity in the
absence of lines.  Therefore we write down immediately the emergent
intensity of a constant $p$ ray originating at the back\footnote{In
  this discussion ``front'' and ``back'' refer to locations in the SN
  nearest to and farthest from the observer, respectively.} of the
core and traversing toward the observer, without calculating the
source function. When no lines are present the intensity of a ray
passing through the core is proportional to its geometric length:
\begin{equation}
  \label{eq:conti}
  I(p)=2j(\rc^2-p^2)^{1/2}.
\end{equation}
A representative intensity ray is shown in
Figure~\ref{fig:core_single_ray_with_z_axes}.
If instead the emissivity has spatial dependence, $j = j(r)$, the
result has a more general form:
\begin{equation}
  \label{eq:cont_with_j_of_r}
  I(p) = \int_{-(\rc^2-p^2)^{1/2}}^{(\rc^2-p^2)^{1/2}} j((p^2 +
  z^2)^{1/2}) dz = 2 \int_0^{(\rc^2-p^2)^{1/2}} j((p^2+z^2)^{1/2}) dz
\end{equation}
where we have transformed the argument of $j$ since the intensity
along a ray of constant $p$ depends only on its $z$ coordinate.

Already a physical peculiarity arises: limb darkening in the absence
of scattering or absorption. The traditional interpretation of limb
darkening in photospheric objects such as dwarf stars and young SNe is
that intensity rays with large impact parameter $p$, that is, rays
which emerge from the limb, reach an optical depth of $\tau \simeq 1$
at shallower layers in the atmosphere than low-impact parameter rays.
Under the assumption of local thermodynamic equilibrium (LTE),
shallower locations in atmospheres have lower temperatures and thus
weaker source functions, since LTE requires by definition that $S = B$
where $B$ is the Planck function. A weaker source function in turn
leads to a lower intensity emergent ray, the cumulative result of
which is limb darkening.

In the case of this post-photospheric model, however, high-$p$ rays
accumulate less continuum as they proceed from the back of the object
toward the observer. Such accumulation does not occur in
photospheric-phase atmospheres of SNe except through lines in the
atmosphere which scatter continuum photons from the blackbody
photosphere into the ray; however, this contribution is small except
when the line has an extraordinarily high optical depth.

Using Equation~\ref{eq:conti}, the continuum flux at all wavelengths is
given by
\begin{equation}
  \label{eq:flux}
  F_\lambda \equiv \int I_\lambda \cos \theta d\Omega =
  \frac{2\pi}{\rc^2} \int_0^{\rc} I_\lambda p dp = \frac{4\pi}{3}j\rc.
\end{equation}
For clarity, and without loss of generality, hereafter we set $j\equiv
1$, so $I_\lambda$, $J_\lambda$, and $F_\lambda$ have units of length
and the geometric nature of our results becomes apparent.  In
addition, since in homologous expansion $v\propto r$, we interchange
lengths and velocities where convenient.

\subsection{Mean intensity}

We now calculate the mean intensity, $J$, both inside and outside the
glowing core. Without any lines, $J$ is not a particularly useful
quantity in this model since we do not need it to calculate the
emergent flux.  However, when we add a single line to the core in
\S\ref{sec:line_in_core} we require the source function to calculate
the emergent intensity, and in the resonance-scattering approximation
$S = J$.

\begin{figure}
\centering
  \includegraphics[scale=0.5]{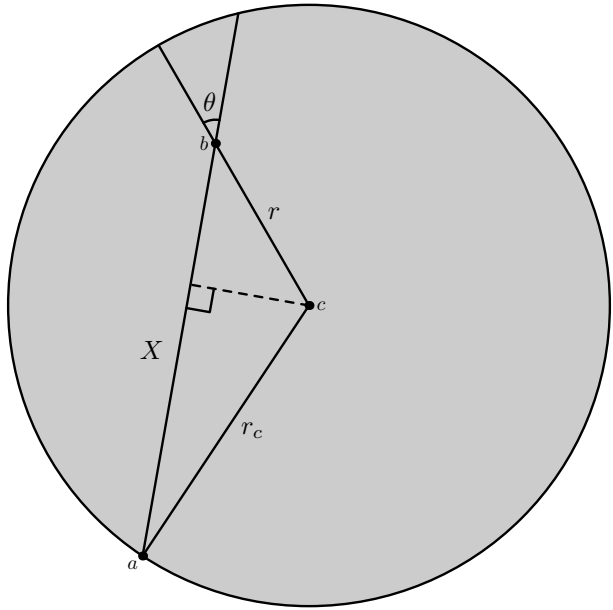}
  \caption{Geometric configuration used to calculate $J(r < \rc)$ both
    for pure continuum and for a single line in the core. $X$ is the
    magnitude of the vector $\protect\overrightarrow{ab}$, $\rc$ is
    that of $\protect\overrightarrow{ac}$, and $r$ is that of
    $\protect\overrightarrow{bc}$.}
  \label{fig:SF_inside_core}
\end{figure}

By definition,
\begin{equation}
  \label{eq:jr}
  J_\lambda(r) \equiv \frac{1}{4 \pi} \int I_\lambda(r) d\Omega =
  \frac{1}{2} \int_{-1}^1 I_\lambda(r) d\mu
\end{equation}
where $\mu \equiv \cos \theta$, and we have applied to the radiation
field the condition of azimuthal symmetry.  Inside the core, $r \leq
\rc$, so from Figure~\ref{fig:SF_inside_core},
\begin{equation}
    X = r \mu + \left(r^2 \mu^2 + \rc^2 - r^2 \right)^{1/2}.
\end{equation}
Setting $I_\lambda(r) = X(r)$, plugging this into
Equation~\ref{eq:jr}, and using standard integral tables, we find
\begin{equation}
  \label{eq:jrin}
  J(r) = \frac{1}{2r}\left\{ r\rc + \frac{(\rc^2 - r^2)}{2} \ln
  \left[\frac{\rc+r}{\rc-r}\right]\right\}.
\end{equation}
Calculating the mean intensity outside the core (with the line still
inside the core) is slightly more complicated and we include the
derivation in Appendix~\ref{apx:S_outside}.  The result is:
\begin{equation}
  J(r) = \frac{1}{2r}\left\{ r\rc + \frac{(r^2-\rc^2)}{2}
  \ln\left[\frac{r-\rc}{r+\rc}\right] \right\}.
  \label{eq:jrout}
\end{equation}

\begin{figure}
\centering
  \includegraphics[scale=0.65]{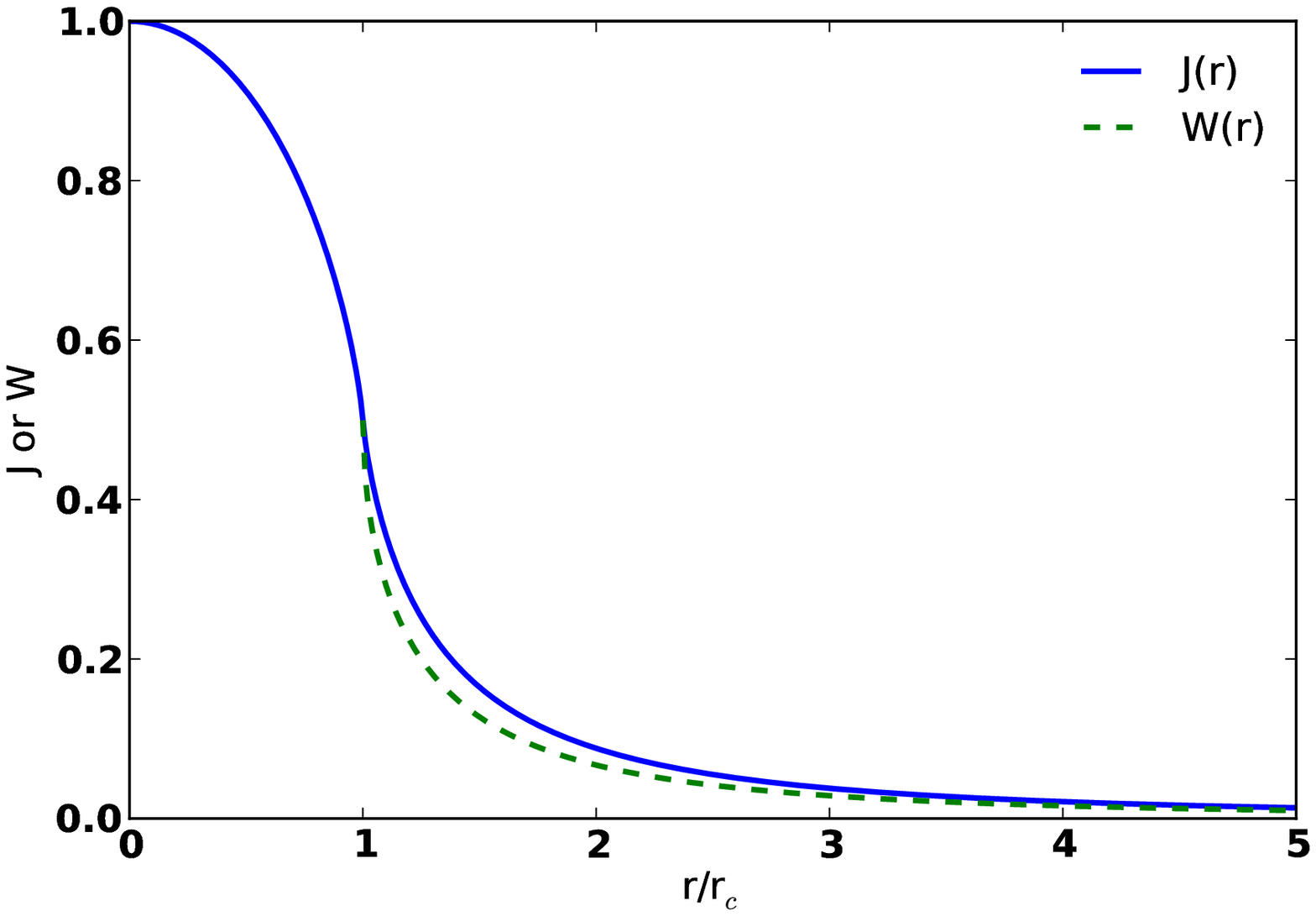}
  \caption{The profile of $J(r)$ and $W(r)$, where $W$ is the dilution
    factor.}
  \label{fig:J_W_plot}
\end{figure}

Figure~\ref{fig:J_W_plot} shows the behavior of $J(r)$ in units of \rc
$\text{}$ and for comparison the dilution factor $W(r)$
\citep{mihalas78sa} is also shown.  The shape of $J$ is vary similar
to that of a Gaussian and it is larger than $W(r)$ until quite large
$r$, where both functions behave as $\frac{1}{4}(\frac{\rc}{r})^2$. We
emphasize that $J$ peaks strongly at $r=0$, a result which affects
line formation significantly, as we will discuss in
\S\ref{sec:line_in_core}.

\section{A line in the core}
\label{sec:line_in_core}

\begin{figure}
\centering \includegraphics[scale=0.45]{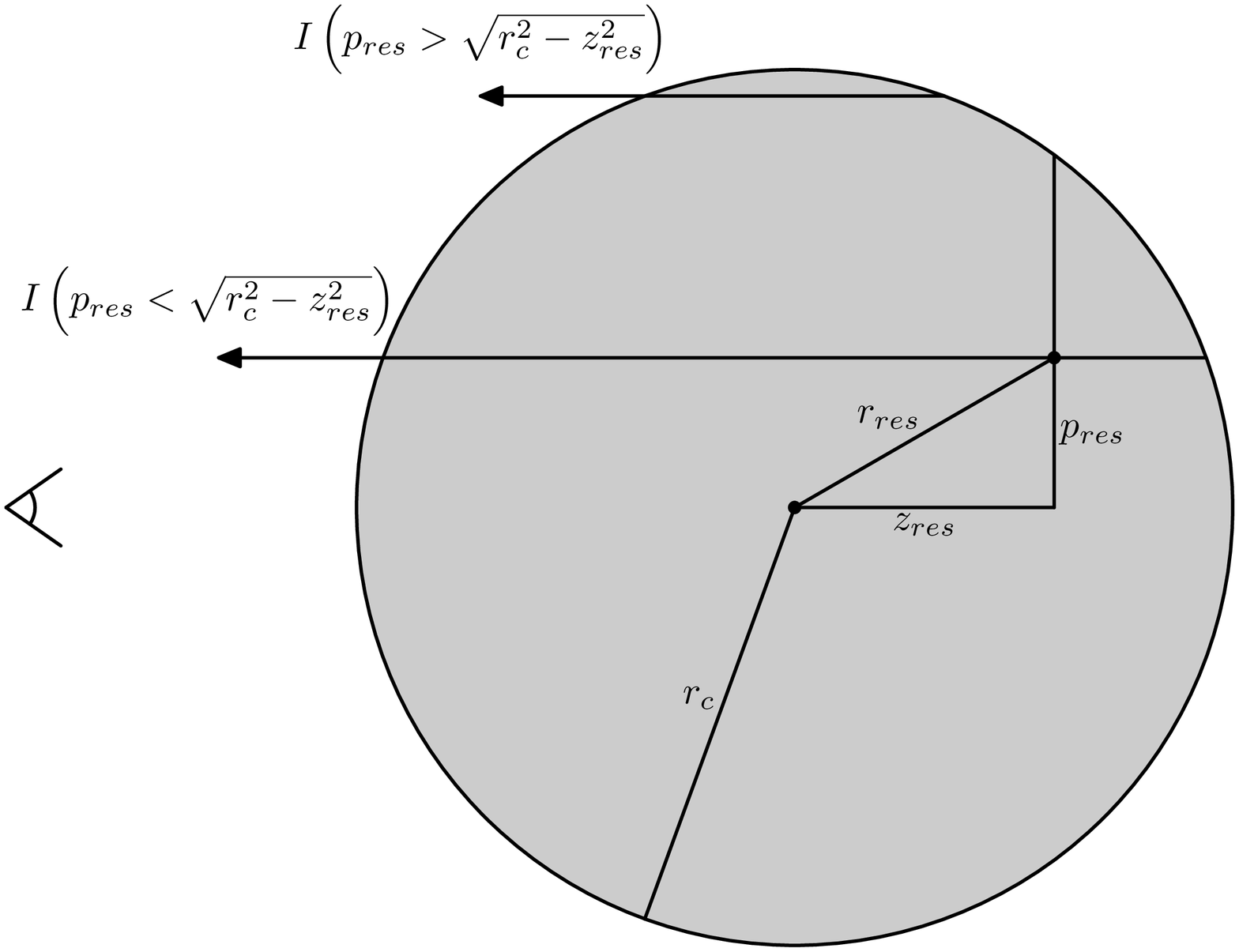}
  \caption{Intensity rays with a single line in the core.}
  \label{fig:core_rays}
\end{figure}

\begin{figure}[ht]
  \centering \includegraphics[scale=0.6]{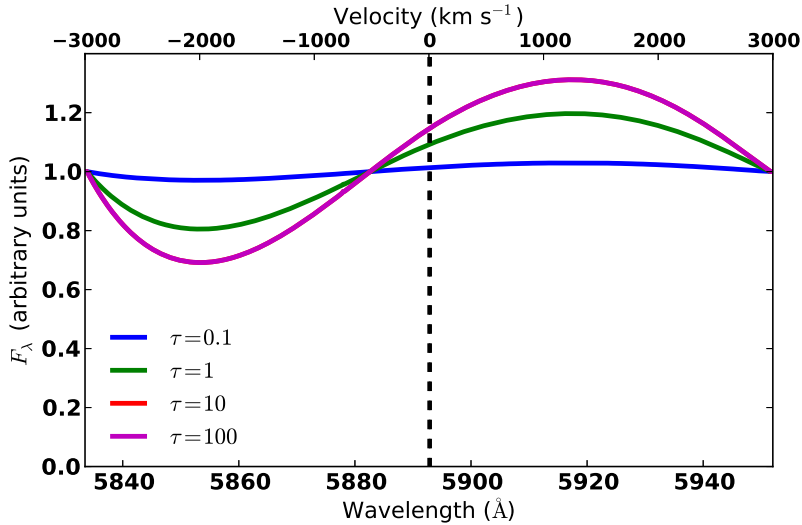}
  \caption{Flux profiles for Na I D, $\lambda 5892$, in the
    post-photospheric model, with spatially constant optical depth and
    an outer core velocity of 3000 \kmps.  The vertical dashed line
    indicates the rest wavelength of the line.  The $\tau = 10$ and
    $\tau = 100$ profiles overlap almost exactly and are
    indistinguishable in this figure. In contrast to \synow the red
    emission peak does not occur at the rest wavelength, but rather is
    redshifted by an amount independent of optical depth.}
  \label{fig:const_tau_one_line}
\end{figure}

\begin{figure}[ht]
  \centering \includegraphics[scale=0.6]{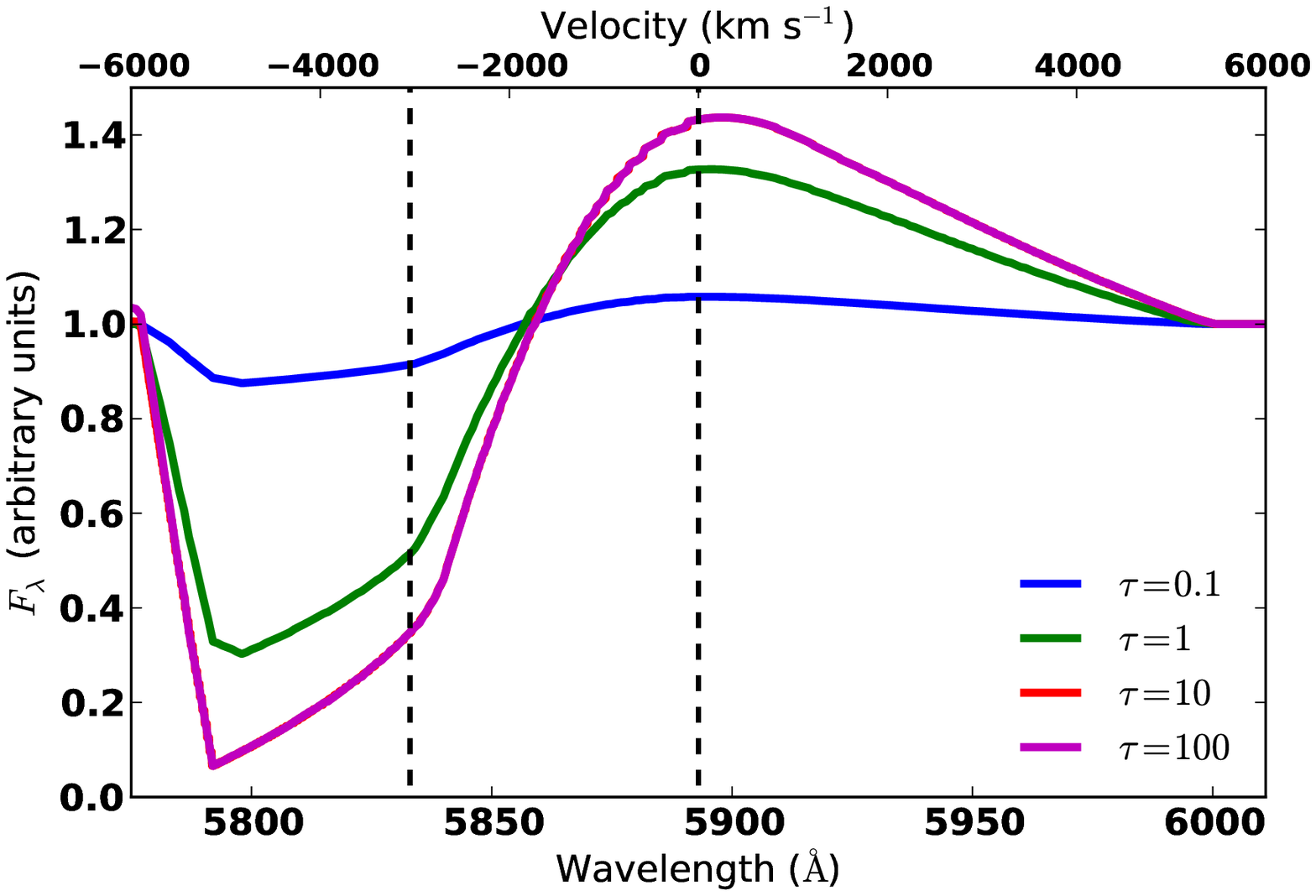}
  \caption{Flux profiles for Na I D in \synow, with photospheric
    velocity 3000 \kmps and spatially constant optical depth. The
    maximum velocity of the line-forming region is 6000 \kmps. The
    right dashed vertical line indicates the rest wavelength of the Na
    I D line ($\lambda 5892$), and the left vertical line indicates
    the blueshifted rest wavelength of Na I D at the photospheric
    velocity. The $\tau = 10$ and $\tau = 100$ profiles overlap almost
    exactly and are indistinguishable in this figure. Unlike the
    profiles in Figure~\ref{fig:const_tau_one_line}, the emission peak
    in the \synow case remains fixed at the line rest wavelength
    regardless of the optical depth.}
  \label{fig:NaID_synow}
\end{figure}

We now treat the case of a single line, forming inside the core. We
assume, for simplicity, a constant Sobolev optical depth $\tau$ in the
line throughout the core.  Before continuing, we note that the most
profound effect of homologously expanding matter on the radiation
field is that photons redshift with respect to the matter regardless
of propagation direction.  Therefore photons in a ray which starts at
the back of the core and emerges toward the observer redshift
continuously as they move along the ray.  Referring to
Figure~\ref{fig:core_rays}, if a ray originates in front of the plane
of constant $\zres$ corresponding to the location in the core $z =
\zres$ where the line Doppler shifts into resonance with a particular
wavelength point --- that is, if $p>(\rc^2-\zres^2)^{1/2}$ --- then
the intensity of that ray is simply its continuum value, given by
Equation~\ref{eq:conti}. However, if the ray forms behind that plane
--- if $p<(\rc^2-\zres^2)^{1/2}$ --- the line attenuates some of the
continuum intensity by scattering photons out of the ray and therefore
out of the observer's line of sight. In the Sobolev approximation this
attenuation manifests as a $e^{-\tau}$ term multiplying the continuum
intensity at the location of the line. In addition to attenuating the
intensity along a ray, the line also contributes to the intensity via
its source function $S(r[\zres])$. Specifically, the contribution is
\begin{equation}
  \label{eq:I_line}
  I_{\mathrm{line}} = S(1-e^{-\tau}).
\end{equation}
The two different types of rays are depicted in
Figure~\ref{fig:core_rays}.
Their emergent intensities are
\begin{equation}
  \label{eq:coreline}
  I_{\zres}(p) = \left\{ \begin{array}{ll} ((\rc^2-p^2)^{1/2}-\zres)
    e^{-\tau} + S(r)(1-e^{-\tau}) + (\rc^2-p^2)^{1/2}+\zres \quad &p <
    (\rc^2 -\zres^2)^{1/2} \\ 2 (\rc^2-p^2)^{1/2} \quad &p > (\rc^2 -
    \zres^2)^{1/2}
  \end{array} \right.
\end{equation}
The integral over $I_{\zres}(p)$ is straightforward except for the
term containing the source function $S(r)$. To integrate this term we define
\begin{equation}
g(\zres) \equiv \int_0^{\sqrt{\rc^2-\zres^2}} S(r) p dp =
\int_0^{\sqrt{\rc^2-\zres^2}} \frac{1}{2r} \left[ r \rc +
  \frac{\rc^2-r^2}{2} \ln \left[ \frac{\rc+r}{\rc-r} \right] \right] p dp.
\end{equation}
We then transform the integration variable from $p$ to $r$, which leads to
\begin{equation*}
g(\zres) = \int_{\zres}^{\rc} \frac{1}{2r} \left[ r \rc +
  \frac{\rc^2-r^2}{2} \ln \left[ \frac{\rc+r}{\rc-r} \right] \right] r
dr.
\end{equation*}
This integral is still unwieldy, so we transform integration variables
once again by defining
\begin{equation*}
\mu \equiv \frac{\zres}{r}
\end{equation*}
and
\begin{equation*}
\mures \equiv \frac{\zres}{\rc}
\end{equation*}
from which we find
\begin{equation*}
\frac{r}{\rc} = \frac{\mures}{\mu}.
\end{equation*}
We then change the integration variable from $r$ to $\mu$, which yields
\begin{equation*}
g(\zres) = \frac{\rc^3}{2} \mures^2 \int_{\mures}^1 \left[ 1 +
    \frac{\mu^2 - \mures^2}{2 \mu \mures} \ln \left[ \frac{1 +
        \mures}{1 - \mures} \right] \right] \frac{d\mu}{\mu^3}.
\end{equation*}
The result is
\begin{equation}
g(\zres) = \frac{\rc^3}{6} \left[ 1 + 2 \ln 2 - \mures^2 -
  \frac{\mures}{2} (3 - \mures^2) \ln \left[ \frac{1 + \mures}{1 -
      \mures} \right] - \ln(1 - \mures^2) \right].
\end{equation}
The remainder of Equation~\ref{eq:flux} poses little challenge and
leads directly to an analytic result for the flux profile for the
constant-$\tau$ case:
\begin{align}
\label{eq:f_z_const_tau}
F(\zres) = \frac{2\pi}{\rc^2} \left[ \frac{2\rc^3}{3} + (1 - e^{-\tau})
  \left[ g(\zres) + \frac{\zres(\rc^2 - \zres^2)}{2} - \frac{\rc^3 -
      |\zres|^3}{3} \right] \right].
\end{align}
We remark here that, although Equation~\ref{eq:flux} defines the flux
as a function of wavelength, we have derived
Equation~\ref{eq:f_z_const_tau} in terms of the coordinate
$\zres$. One transforms between $z \leftrightarrow \lambda$ using the
first-order Doppler formula,
\begin{equation}
\label{eq:doppler}
z = r_{\mathrm{max}} \left(\frac{c}{v_{\mathrm{max}}}\right)
\frac{\lambda - \lambda_0}{\lambda_0},
\end{equation}
where $r_{\mathrm{max}}$ ($v_{\mathrm{max}}$) is the maximum radius
(velocity) of the ejecta. We show in
Figure~\ref{fig:const_tau_one_line} the line profiles of Na I D,
$\lambda 5892$, with several different optical
depths.\footnote{Examining Figures~\ref{fig:synow_03du} and
  \ref{fig:synow_94D}, identification of Na I is ubiquitous; we
  therefore use the Na I D line for illustrative purposes.} For
comparison we show in Figure~\ref{fig:NaID_synow} the profile of Na I
D as calculated by \synow, with the same optical depths as in
Figure~\ref{fig:const_tau_one_line}.

The profiles in both plots appear quite P~Cygni-like, but differences
do exist between them. We highlight two in particular. First, when a
photosphere is present as in \synow, the emission peak of a line
profile is located at its rest wavelength, regardless of the velocity
of the material forming the line in the SN atmosphere. This fact
simplifies enormously the task of identifying isolated lines in a SN
spectrum \citep{jb90}. In the post-photospheric model, however, the
emission peak is \emph{not} centered on the line rest wavelength;
rather it is considerably redshifted. As
Figure~\ref{fig:const_tau_one_line} shows, with a core outer edge
velocity of only 3000~km~s$^{-1}$, the emission peak is $\sim 15$
\AA\ redward of the line rest wavelength. Though the absorption
component of a P~Cygni profile is used more often than the
corresponding emission to identify lines in a spectrum \citep[since
  the former dominates the latter in overlapping lines; see][]{jb90},
this redshift could lead to misidentification of lines in
intermediate- or late-time SN spectra.

What is the origin of the emission peak redshift? This question is
easier to answer in the limit $\tau \rightarrow \infty$, the line
profile for which would look identical to that of $\tau = 100$ in
Figure~\ref{fig:const_tau_one_line}. In the $\tau \rightarrow \infty$
case the line at the resonance point $\zres$ scatters \emph{all}
continuum photons forming in the ray segment $z > \zres$ out of the
ray, and the contribution of the line to the intensity along that ray
is simply $I_{\mathrm{line}} = S(r[\zres])$. Thus only two sources of
emission contribute to the flux at a given wavelength: 1.)  the
continuum emission in front of the resonance plane at $z = \zres$; and
2.) the source function at the plane. If the plane is located near the
back of the core, $z \lesssim \rc$, the portion of the core's volume
which emits continuum photons which escape unscattered and reach the
observer is large. However, because the surface area of the plane is
small when $z$ is close to $\rc$, and because from
Figure~\ref{fig:J_W_plot} we see that $S$ is small at $r \sim \rc$ ---
only half of its maximum value --- the emission contribution from the
source function on the plane is in turn relatively small.

As the resonance plane moves forward (to bluer wavelengths, smaller
$z$), the volume of the emitting core in front of the plane decreases,
but the surface area of the plane grows, and Figure~\ref{fig:J_W_plot}
shows that the source function at the plane grows quickly as
well. From $\frac{1}{3}\rc\lesssim z \lesssim \rc $ the scattering
emission from the resonance plane more than compensates for the
diminishing continuum emission from the core, causing the flux to
increase monotonically as one moves blueward in that region,
eventually reaching the emission peak, which in
Figure~\ref{fig:const_tau_one_line} is $\sim 5915$ \AA.  Blueward of
this peak, despite the fact that $S$ increases monotonically until
reaching $z = 0$, the surface area of the plane increases only
slightly when $z \gtrsim 0$ and the now very large resonance plane
occults so much of the core over a small shift $\Delta z$ that it can
no longer compensate for the large amount of emission removed from the
volume of the core behind it, causing the flux to decrease as $z$
becomes smaller.  This transition point exists \emph{redward} of $z =
0$ and thus the emission peak of the P~Cygni profile is redder than
the rest wavelength of the line.

Blueward of $z=0$, that is, $-\rc < z < 0$, the area of the resonance
plane begins to decrease, now only obscuring a cylindrical volume of
the glowing core.  Even though the length of this occulted cylinder
increases as $z$ becomes more negative, its radius decreases, and the
portion of the total volume of the core that this cylinder comprises
decreases as well.  The now-unocculted limbs of the core, emitting
continuum which the resonance plane can no longer scatter away, grow
in volume and eventually compensate once again for the cylindrical
volume obscured by the plane, causing the flux to increase.  This
transition point manifests as the absorption minimum in the flux
profile ($\sim 5855$ \AA$\text{}$ in
Figure~\ref{fig:const_tau_one_line}), blueward of which the flux
increases until the resonance plane reaches $z = -\rc$ and we recover
pure continuum.

The second difference between the line profiles with and without a
photosphere is that, in the former, the flux in strong lines deviates
from the continuum by a large amount, whereas in the latter the
changes are small. The flux in the absorption minimum of the $\tau_0 =
100$ line in Figure~\ref{fig:NaID_synow}, for example, is 80\% lower
than the continuum value. In our post-photospheric model, on the other
hand, even the strongest line in Figure~\ref{fig:const_tau_one_line}
departs by only up to 30\% from the continuum. That the
post-photospheric model exhibits such small departures from continuum
is due to the ability of a small portion of the core to emit
unscattered photons toward the observer even when the resonance plane
is near the front. In the \synow case, the shell above the photosphere
does not emit any continuum, so when the plane is close to the front
of the ejecta, only scattering from the plane itself contributes to
the flux, causing it to be extremely low at the absorption minimum.

\begin{figure}[ht]
\centering
  \includegraphics[scale=0.6]{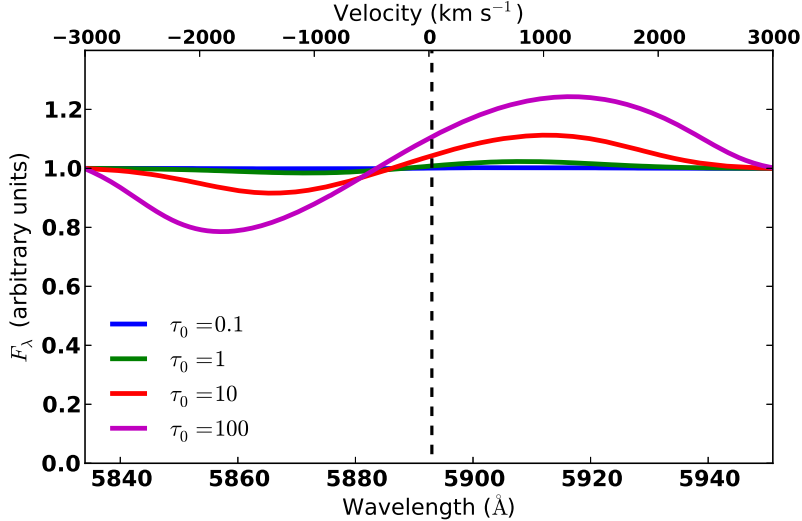}
  \caption{Flux profiles for Na I D, $\lambda 5892$, in the
    post-photospheric model with outer edge core velocity 3000 \kmps
    and a line optical depth following an exponential decay, $\tau(v)
    = \tau_0 \exp(-v / v_e)$, where $v_e = 500$ \kmps. The vertical
    dashed line indicates the rest wavelength of the line. In contrast
    to \synow, the red emission peak does not occur at the rest
    wavelength, but is rather significantly redshifted by an amount
    which depends on the strength of the line. Also unlike \synow, the
    blue absorption minimum blueshifts continuously with increasing
    optical depth.}
  \label{fig:exp_tau_one_line}
\end{figure}

\begin{figure}[ht]
\centering
  \includegraphics[scale=0.6]{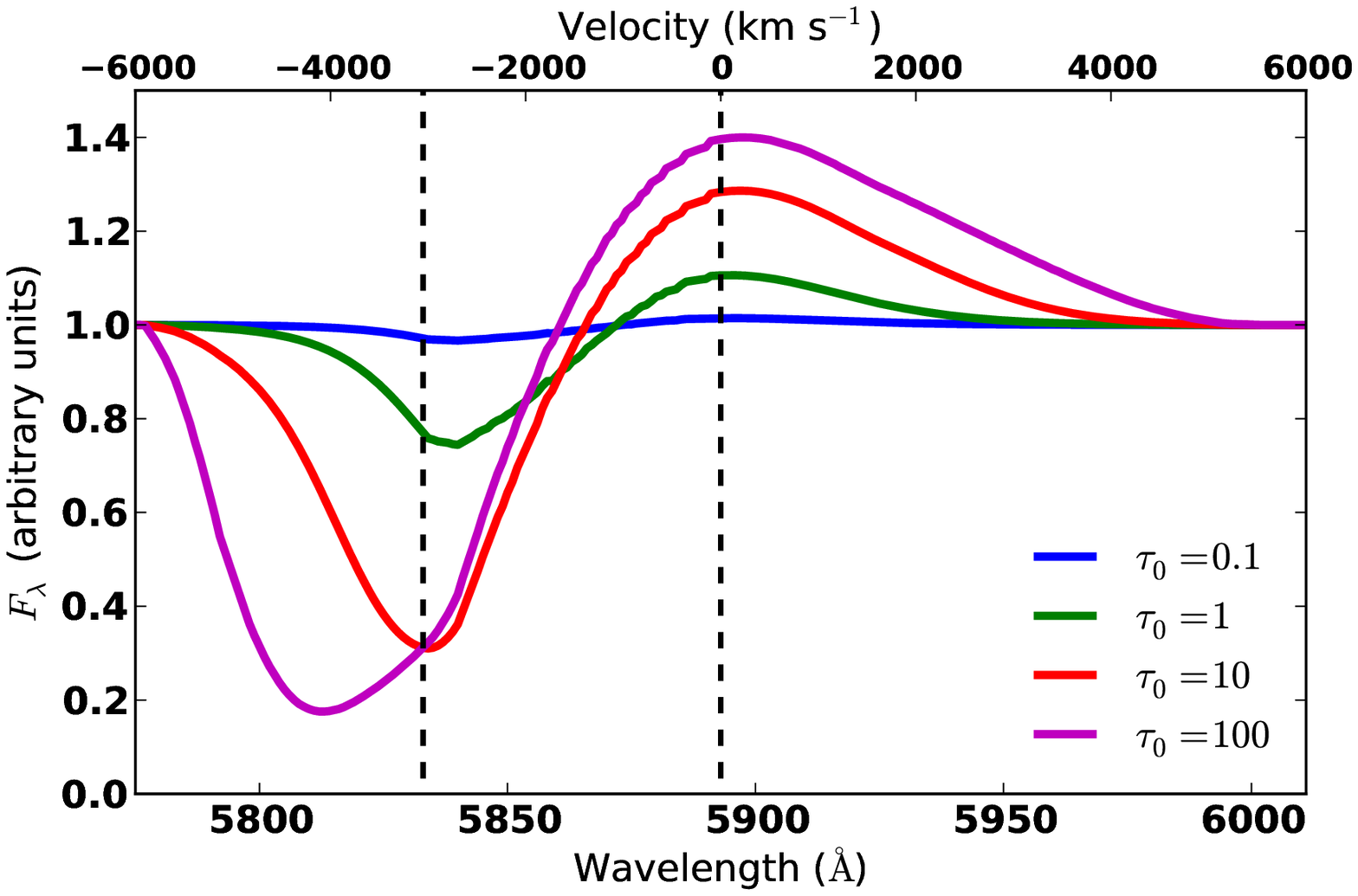}
  \caption{Flux profiles for Na I D, $\lambda 5892$, in \synow, with
    photospheric velocity 3000 \kmps and a line optical depth
    following an exponential decay, $\tau(v) = \tau_0 \exp(-v / v_e)$,
    where $v_e = 500$ \kmps. The maximum velocity of the line-forming
    region is 6000 \kmps. The right vertical dashed line indicates the
    rest wavelength of the line, and the left dashed line indicates
    the wavelength of Na I D blueshifted to the photospheric
    velocity. Except for extremely high optical depths, the blue edge
    of the absorption minimum remains fixed at the photospheric
    velocity; and for all optical depths, the red emission peak
    remains at the rest wavelength of the line.}
  \label{fig:synow_exp_tau_one_line}
\end{figure}

The physical simplicity of constant $\tau$ line profiles such as those
in Figures~\ref{fig:const_tau_one_line} and \ref{fig:NaID_synow}
readily facilitates analysis of line formation in SN ejecta, as we
have just seen. However, one must also be aware of complicating
effects such as the inhomogeneous structure of SNe, viz. the
velocity-dependent density profile. In the Elementary Supernova
framework one assumes implicitly that $\tau = \tau(\rho)$, where
$\rho$ is the matter density, and accounts for this by writing $\tau$
as, e.g., an exponential or power law function which decreases with
increasing velocity. To illustrate how a variable optical depth
affects line formation we show in Figure~\ref{fig:exp_tau_one_line}
the line profile of Na I D in our post-photospheric model with
$\tau(v) = \tau_0 \exp(-v/v_e)$, where $\tau_0$ is a constant and $v_e
= 500$ \kmps. Again, for comparison, we show in
Figure~\ref{fig:synow_exp_tau_one_line} a \synow plot of the same line
with the same exponential $\tau$ profile.

In the post-photospheric case, shown in
Figure~\ref{fig:exp_tau_one_line}, the relationship between velocities
and the positions of the emission maximum and absorption minimum is
not immediately obvious. Like the \synow line profiles, shown in
Figure~\ref{fig:synow_exp_tau_one_line}, the blue edge of the
absorption component (not of the absorption \emph{minimum}; see
5835~\AA\ in Figure~\ref{fig:exp_tau_one_line}, and 5775~\AA\ in
Figure~\ref{fig:synow_exp_tau_one_line}) indicates the maximum
velocity of the line-forming material, although in blended cases this
can be difficult to discern. Unlike \synow, in this model both the
peak and trough continuously redshift and blueshift, respectively, as
the line optical depth increases. In the \synow case, on the other
hand, the separation between the rest wavelength of the line and the
minimum of the blueshifted absorption component indicates the velocity
of the photosphere and thus the minimum velocity of the line-forming
material itself \citep[assuming the material is not ``detached'' above
  the photosphere,][]{jb90}. Only when the line optical depth is
extremely high does the location of the absorption minimum begin to
move blueward of the photospheric blueshift, as in the $\tau_0 = 100$
line in Figure~\ref{fig:synow_exp_tau_one_line}. For each of the three
weaker lines in that figure, one can simply measure the blue edge of
the absorption minimum, $\sim~5830$~\AA, then calculate the
photospheric velocity, $\sim~3000$~\kmps.

We also note that the $\tau = 10$ and $\tau = 100$ line profiles in
Figures~\ref{fig:const_tau_one_line} and \ref{fig:NaID_synow} appear
identical, whereas in Figures~\ref{fig:exp_tau_one_line} and
\ref{fig:synow_exp_tau_one_line} the $\tau_0 = 10$ and $\tau_0 = 100$
profiles are distinct. This difference is due to the velocity
dependence of $\tau$: in the former pair of figures, the optical depth
is constant everywhere, leading to a saturated profile throughout the
line-forming region when $\tau \geq 10$. In the latter pair, on the
other hand, $\tau$ decreases exponentially with increasing velocity,
so the line samples such large optical depths at only very low
velocities. This velocity dependence affects line formation outside
the core in exactly the same way (cf. \S
\ref{sec:one_line_outside_core}).

\section{A line outside the core}
\label{sec:one_line_outside_core}
Having explored in detail the geometric effects of continuum emission
on line formation in the core, we now introduce a non-emitting,
transparent shell around the core, with outer radius $\rmax$, where
$\rmax > \rc$.  In this region there is no continuum emission and line
formation occurs in exactly the same way as in \synow.  Therefore,
unlike the case discussed in \S\ref{sec:line_in_core}, the line
forming region now exists \emph{outside} the core, rather than inside.
We include the shell in this model to account for the possibility of
intermediate mass elements such as Ca II forming lines at late times
above the material which has been burned all the way to the iron-peak.

\subsection{Intensity for a line outside the core}
\label{apx:intensity_in_shell}

\begin{figure}
\centering \includegraphics[scale=0.45]{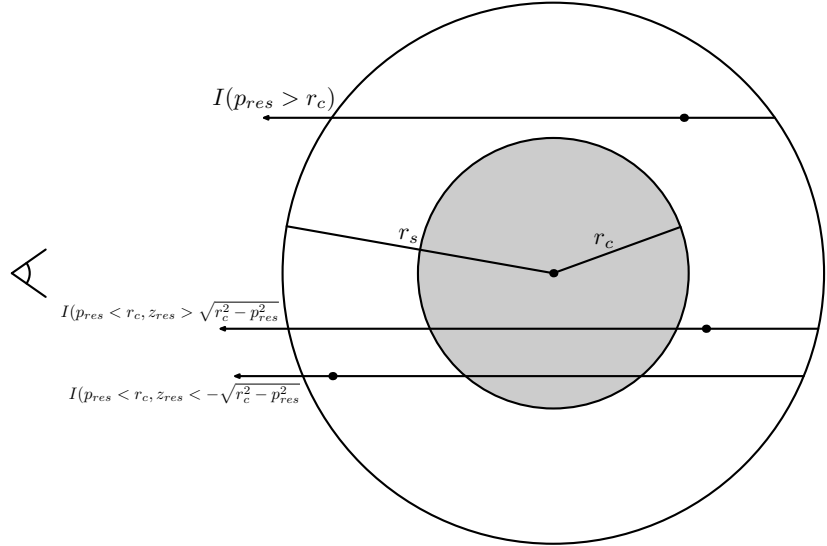}
  \caption{The three types of intensity rays for the core+shell
    configuration.}
  \label{fig:shell_rays}
\end{figure}

We must now consider a number of possible ways that the line and a
particular ray can interact.  First, if the ray has impact parameter
$p > \rc$ then it never intersects the core and, regardless of the
location of the line resonance point $\zres$, the ray's emergent
intensity is
\begin{equation}
  I_{\zres}(p)=S(r[\zres])(1-e^{-\tau})
\label{eqn:I_out1}
\end{equation}
where $S(r)$ is now the source function \emph{outside} the core, given
by Equation~\ref{eq:jrout}, and
\begin{equation}
  \tau = \tau(r[\zres]).
\end{equation}
If the line has $p<\rc$ but $\zres < -\sqrt{\rc^2 -
  p_{\mathrm{res}}^2}$, then the line attenuates the continuum ray
from the core:

\begin{equation}
  I_{\zres}(p) = S(r[\zres])(1-e^{-\tau})+2(\rc^2-p^2)^{1/2}e^{-\tau}.
\label{eqn:I_out2}
\end{equation}
Finally, if the line has $p < \rc$ and $\zres > \sqrt{\rc^2 -
  p_{\mathrm{res}}^2}$, then it does \emph{not} attenuate the core
continuum and the emergent intensity is
\begin{equation}
  I_{\zres}(p) = S(r[\zres])(1-e^{-\tau})+2(\rc^2-p^2)^{1/2}.
\label{eqn:I_out3}
\end{equation}
These three cases are shown in Figure~\ref{fig:shell_rays}.

\subsection{Flux for a line outside the core}
\label{sec:flux_in_shell}

\begin{figure}
\centering \includegraphics[scale=0.4]{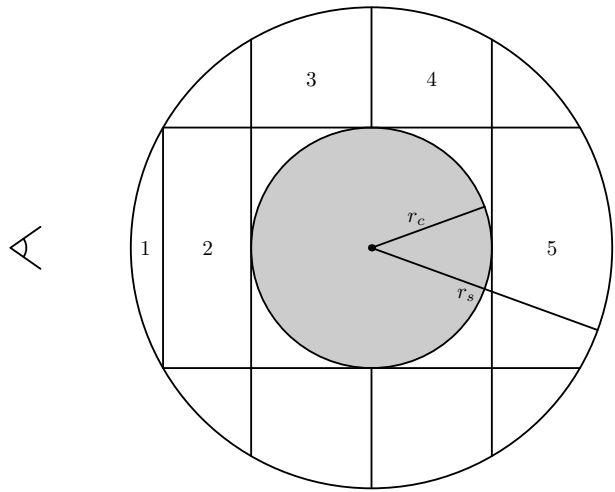}
  \caption{The five flux regions in the core+shell configuration.}
  \label{fig:shell_flux_zones}
\end{figure}

While $F_\lambda$ proceeds simply from $I_\lambda$ in the case $r \leq
\rc$, its form is much more complicated in the $r > \rc$ regime. In
particular the flux integral takes a unique form in five different
regions. We show the five zones in Figure~\ref{fig:shell_flux_zones}.

\begin{enumerate}
\item $-r_s < z < -(r_s^2-\rc^2)^{1/2}$
  \begin{eqnarray*}
    F_z &=& \frac{4 \pi}{\rc^2} e^{-\tau} \int_0^{(r_s^2-z^2)^{1/2}}
    (\rc^2-p^2)^{1/2} p dp \\ &&+ \frac{4\pi}{\rc^2}
    \int_{(r_s^2-z^2)^{1/2}}^{\rc} (\rc^2-p^2)^{1/2} p dp \\ &&+ \frac{2
      \pi}{\rc^2} (1-e^{-\tau}) \int_0^{(r_s^2-z^2)^{1/2}} S(r[p]) p dp
  \end{eqnarray*}
\item $-(r_s^2-\rc^2)^{1/2} < z < -\rc$
  \begin{equation*}
    F_z = \frac{4\pi}{3}\rc e^{-\tau} + \frac{2\pi}{\rc^2}
    (1-e^{-\tau}) \int_0^{(r_s^2-z^2)^{1/2}} S(r[p]) p dp
  \end{equation*}
\item $-\rc < z < 0$
  \begin{eqnarray*}
    F_z &=& \frac{4\pi}{\rc^2} \int_0^{(\rc^2-z^2)^{1/2}}
    (\rc^2-p^2)^{1/2} p dp \\ &&+ \frac{4\pi}{\rc^2} e^{-\tau}
    \int_{(\rc^2-z^2)^{1/2}}^{\rc} (\rc^2-p^2)^{1/2} p dp \\ &&+
    \frac{2\pi}{\rc^2}(1-e^{-\tau})
    \int_{(\rc^2-z^2)^{1/2}}^{(r_s^2-z^2)^{1/2}} S(r[p]) p dp
  \end{eqnarray*}
\item $0 < z < \rc$
  \begin{equation*}
    F_z = \frac{4\pi}{3}\rc + \frac{2\pi}{\rc^2}(1-e^{-\tau})
    \int_{(\rc^2-z^2)^{1/2}}^{(r_s^2-z^2)^{1/2}} S(r[p]) p dp
  \end{equation*}
\item $\rc < z < r_s$
  \begin{equation*}
    F_z = \frac{4\pi}{3}\rc + \frac{2\pi}{\rc^2}(1-e^{-\tau})
    \int_0^{(r_s^2-z^2)^{1/2}} S(r[p]) p dp
  \end{equation*}
\end{enumerate}

\begin{figure}
\centering
\includegraphics[scale=0.6]{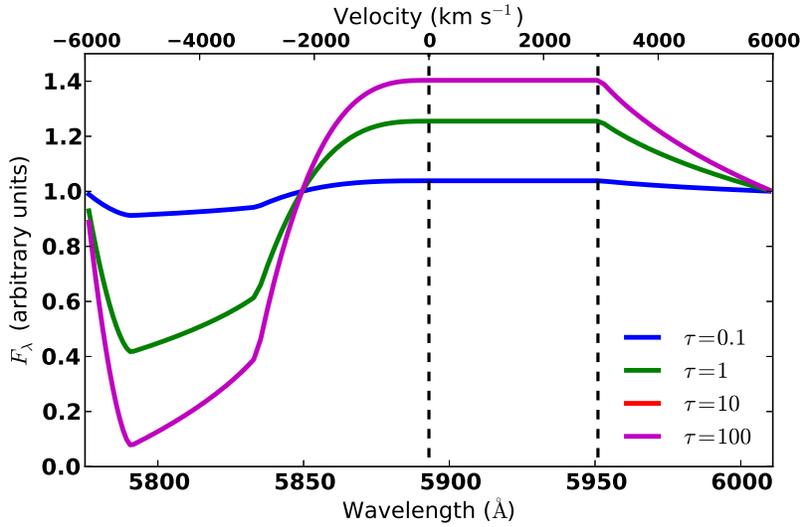}
\caption{Flux for Na I D in the shell, with core velocity 3000 \kmps
  and outer edge shell velocity 6000 \kmps. The rest wavelength of the
  line is $\lambda_0 = 5892$ \AA, indicated by the left vertical
  dashed line. The right dashed line indicates the rest wavelength of
  Na I D redshifted to the back edge of the core, at which point the
  emission peak forms a plateau. For $\rc \leq r \leq \rmax$, $\tau =
  \mathrm{constant}$ and for $r \leq \rc$, $\tau = 0$. The $\tau = 10$
  and $\tau = 100$ profiles overlap almost exactly and are
  indistinguishable in this figure.}
\label{fig:flux_outside_core}
\end{figure}

Figure~\ref{fig:flux_outside_core} depicts single-line profiles with
different values of $\tau$. We first note the resemblance of these
line profiles to those of \synow shown in
Figure~\ref{fig:NaID_synow}. This is not surprising, given that the
only difference between the two models is that in \synow the core is
opaque while in the post-photospheric model is it transparent. Both
exhibit strong deviations from the continuum flux value, especially in
the absorption component. We discussed in \S \ref{sec:line_in_core}
the cause of this near-zero flux in the absorption component of the
line.

There is one major difference between the line profiles in
Figures~\ref{fig:NaID_synow} and \ref{fig:flux_outside_core}: the
flat-topped emission component in the latter. To discern the source of
this plateau in the spectrum, we conduct a geometric analysis similar
to that presented in \S\ref{sec:line_in_core}, again studying the
limit $\tau \rightarrow \infty$. Starting with the resonance plane at
the back of the shell, $z=r_s$, we see that as it moves forward toward
the observer, its surface area grows, which leads to the flux increase
from 6010~\AA\ blueward to 5950~\AA\ in
Figure~\ref{fig:flux_outside_core}.

When the resonance plane reaches the back edge of the core and begins
to move forward through it, that is, when $0 < \zres < \rc$, the core
projects onto the plane a central circular region where $\tau = 0$ and
continuum emission from the core is unattenuated. The component of the
plane which samples the optical depth in the shell is a ring with area
\begin{equation}
A = \pi (r_s^2 - \zres^2) - \pi (r_c^2 - \zres^2) = \pi(r_s^2 -
r_c^2).
\end{equation}
From this equation we see that $A$ is constant when $0 < \zres < \rc$;
this constancy is the cause of the emission plateau in
Figure~\ref{fig:flux_outside_core}.

When $-\rc < \zres < 0$, the plane begins to obscure the core,
starting at the core's limb, and scatters an increasing amount of the
continuum emission out of the observer's line of sight. When $\zres <
-\rc$ the entire core is obscured and the only emission from the
ejecta comes from the source function at $\zres$. In the region $-r_s
< \zres < -\rc$, line formation occurs in exactly the same way as in
\synow, leading to the absorption trough in
Figure~\ref{fig:flux_outside_core} which is almost identical to that
of the \synow profile shown in Fig. \ref{fig:NaID_synow}.

The constant-flux emission in the specta in
Figure~\ref{fig:flux_outside_core} is the most distinguishing feature
of our post-photospheric model. In \S\ref{sec:disc} we will consider a
nebular line-forming region with a ``hole'' devoid of line optical
depth, and there we will encounter a geometric conspiracy similar to
the one presented in this section, leading to similarly flat features
in line profiles. We remark in addition that \synow can produce
flat-topped emission features in a spectrum by detaching lines from
the photosphere. Thus we conclude that in general a plateau-shaped
line emission feature indicates some kind of missing line opacity.

\section{Two lines in the core}
\label{sec:two_lines_in_core}
We now study the case of two resonance lines in the
core. Understanding the ways in which multiple lines ``interact'' via
their overlapping flux profiles, and in turn being able to identify
them individually, is critical to interpreting observed spectra since,
in all epochs of a SN, its optical spectrum exhibits severe line
blending.

In the resonance-scattering approximation, for any number of lines in
the core, the source function of the bluest line always has the
single-line form given in Equation~\ref{eq:jrin}, since there are no
bluer photons in the radiation field which can redshift into resonance
with it. The second bluest line then interacts only with its bluer
neighbor; the third bluest line interacts with its two bluer
neighbors; etc. In computational terms this means that one calculates
the source function for each line starting with the bluest and moving
redward.  The details of the calculation are shown in
Appendix~\ref{apx:two_lines}.

\begin{figure}
\centering
\includegraphics[scale=0.8]{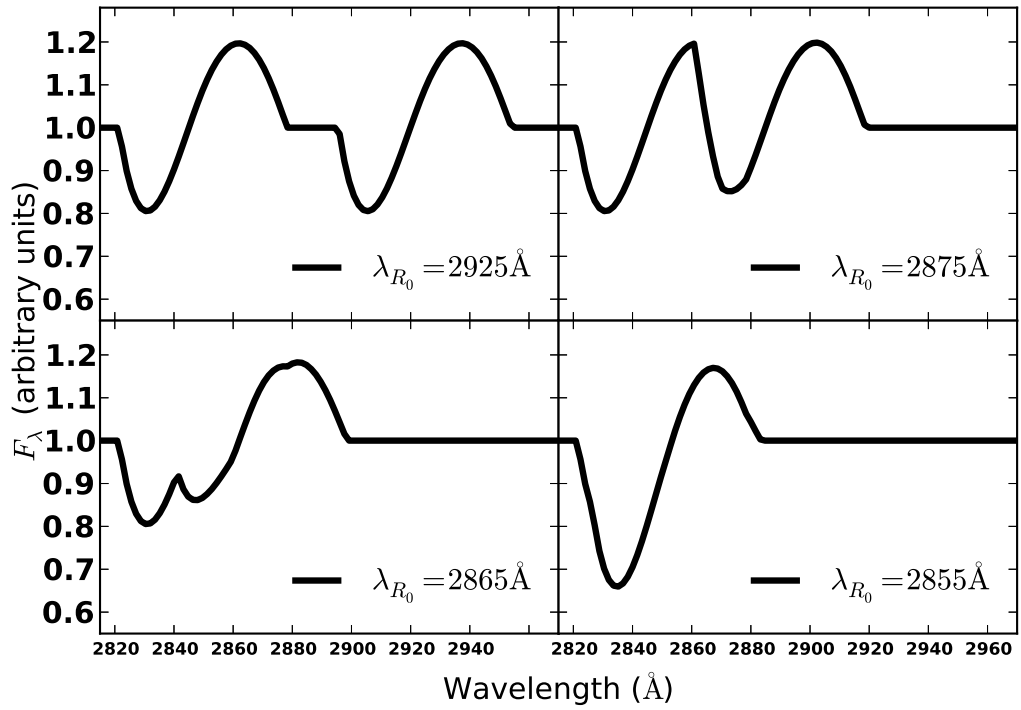}
  \caption{Interaction of two lines in the core as they blend together
    due to Doppler broadening.  The rest wavelength of the blue line
    is fixed at $\lambda_{B_0} = 2850$ \AA, while the rest wavelength
    of the red line moves blueward. The optical depths of both lines
    are fixed at $\tau_B = \tau_R = 1$.}
  \label{fig:two_lines_move_red_line_blueward}
\end{figure}

We first consider the case $v_{\mathrm{core}} = 3000$ \kmps, and we
hold the optical depths of both lines fixed everywhere in the core,
$\tau_B = \tau_R = 1$.  We then set the rest wavelength of the blue
line to 2850~\AA\ and decrement the rest wavelength of the red line
from 3000 \AA, where the lines are too far apart in wavelength space
to blend in the spectrum, to 2855 \AA, where the lines overlap
completely.  The result is shown in
Figure~\ref{fig:two_lines_move_red_line_blueward}.  In this figure we
find that line blending in the post-photospheric case occurs in
essentially the same way as in the \synow case
\citep{jb90}. Specifically, the absorption component of the red line
completely overwhelms the emission of the blue line, and also, when
the two lines overlap perfectly (the bottom-right panel of
Figure~\ref{fig:two_lines_move_red_line_blueward}), the redshifted
emission component appears the same as the single-line case, but the
absorption is much deeper.

One peculiar feature of
Figure~\ref{fig:two_lines_move_red_line_blueward} is that, when the
lines overlap only partially, the emission component of the blue line
acquires a triangular shape. In fact, in the upper-right panel of that
figure, the triangular emission peak is nearly symmetrical.  However
the abrupt (perhaps unrealistic) drop in $\tau$ at the edge of the
core in this calculation may somewhat exaggerate this triangular
feature.

In this section we have restricted our discussion to the interaction
between two lines confined to the core. However, with the introduction
of multiple lines, many other types of interactions are possible. As
an example, in Appendix \ref{apx:two_lines_outside} we construct the
geometric framework for the two-line case where both lines are outside
the core. However the situation is more complex due to the myriad
combinations of line locations as well as their combined interaction
with the core and shell. We thus defer a thorough examination of all
the possible effects that can occur in this case, including the
effects of the photosphere before it has receded completely, to a
future work (B.~Friesen \& R.~C.~Thomas, in preparation).

\section{Discussion}
\label{sec:disc}

Much recent work analyzing \sneia spectra has focused on the
importance of asymmetries \citep{maedanature10, maeda11, maund10a,
  foley-kasen11, foley12b}.  However, the spectra produced by this
model illustrate a potentially significant complication in
interpreting spectra at late times in \sneia, specifically in the
interpretation of line emission in the absence of scattering. To
illustrate this point we refer to the work of \citet{maedanature10,
  maeda11}, who interpret observed blueshifts and redshifts of Fe
lines in nebular spectra as resulting from asymmetric clumps of
iron-peak material moving toward and away from the observer,
respectively, in the optically thin \snia ejecta. This interpretation
implies that the explosion of a white dwarf resulting in a \snia is
itself highly asymmetric, but relies heavily upon the assumption that
spectral features at late times arise \emph{only} through forbidden
transitions of atomic lines due to the low gas density, and that the
resulting emission profiles are distributed roughly symmetrically
about the rest wavelength.  Indeed, this assumption has become
widespread in nebular spectrum synthesis since the pioneering work of
\citet{axelphd80}.  However, if an appreciable amount of
resonance-scattering is present at this epoch, our results show that
in the presence of significant continuum emission, a perfectly
spherically symmetric distribution of matter produces a redshifted
emission component. It remains unclear what effect, if any, that
resonance-scattering has on the sample considered in
\cite{maedanature10} since the objects in their study were often
several hundred days older than the objects we attempted to fit using
\synow in Figures~\ref{fig:synow_03du} and \ref{fig:synow_94D}.

\citet{maund10a} studied polarization measurements of the objects in
\citet{maedanature10} for which such data were available, and drew
similar conclusions as \cite{maedanature10}, that is, that asymmetry
alone may explain the diversity observed in SNe Ia. The survey in
\citet{maund10a} was biased in that polarization data existed only for
SNe with redshifted emission peaks for particular lines of Fe II and
Ni II, but nevertheless we note that, at intermediate times, for many
features there are multiple possible line identifications, and
redshifted emission can result from either optically thin, receding
ejecta, or, as shown in Figure~\ref{fig:const_tau_one_line}, from
spherically symmetric ejecta with large line optical depth. While the
very late data \citep{maedanature10,maund10a} may indeed be showing
the inferred asymmetry, our results show that one should indeed be
cautious about the epoch of the data when interpreting redshifted
emission profiles.

Finally, we consider another significant effect of
resonance-scattering at late times in SNe. Studies of late-time line
profiles and the imaging of the supernova remnant S Andromeda
\citep{hof03du04,moto06,fesen07,gerardy07,maedanature10} have lead to
the suggestion that a ``nickel hole'' exists in at least some
\sneia. In optically thin media, a central region devoid of
line-forming material manifests in a spectrum as a flat-topped
emission feature in the spectrum; \citet{hof03du04} find just this
when studying the [Fe~II] $\lambda 16,440$ line in a spectrum of SN
2003du taken $\sim~300$ days after explosion, suggesting that
$^{56}$Fe and therefore $^{56}$Ni were absent in the central part of
the SN ejecta. However, the IR observations in these studies are quite
noisy, and if the same effect could be observed earlier, or in
stronger lines, there would be more flux, allowing statistics to be
built up on the existence of a nickel hole in \sneia, which would
provide important constraints on the underlying explosion model.

\begin{figure}
\centering
\includegraphics[scale=0.85]{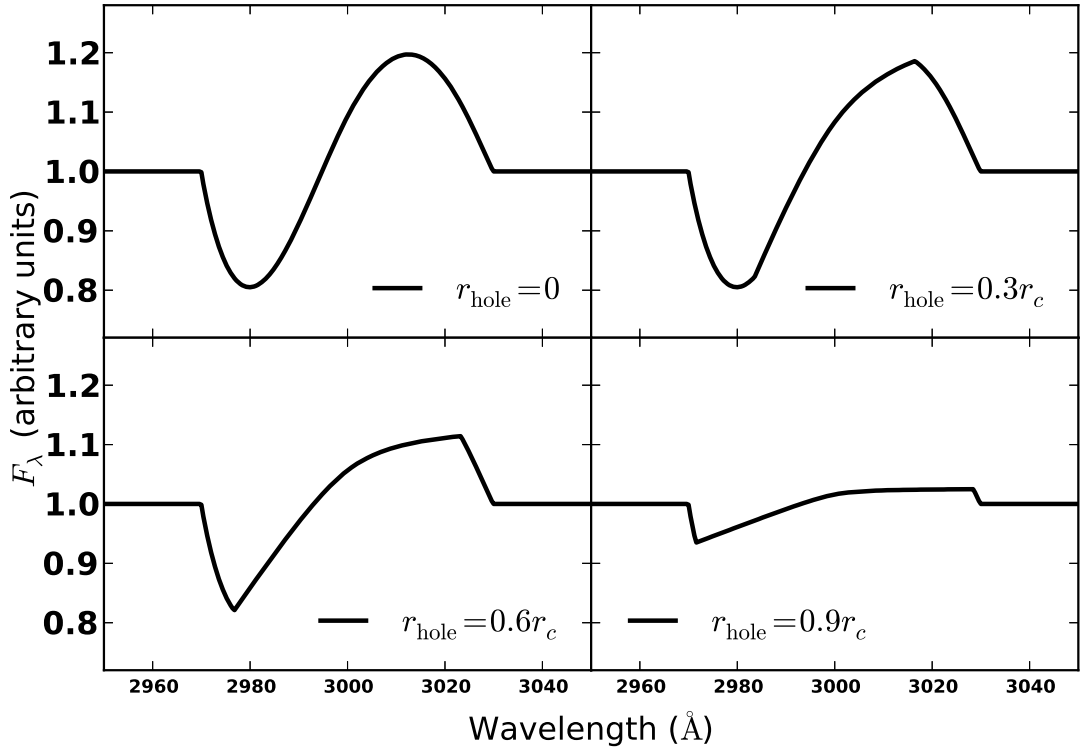}
\caption{A post-photospheric line profile where the line is excluded
  from the various parts of the inner region of the core. The core has
  outer velocity 3000 \kmps. The optical depth in the region $r >
  r_{\mathrm{hole}}$ is $\tau = 1$, whereas in $r <
  r_{\mathrm{hole}}$, $\tau = 0$. The rounded profile is replaced by a
  flattened profile, but there is still an emission peak and an
  absorption trough, both of which become more muted as
  $r_{\mathrm{hole}}$ increases.}
\label{fig:nickel_hole}
\end{figure}

With this in mind we explore the possibility that one may observe a
similar plateau feature inside the glowing core by carving out regions
of zero optical depth in a resonance-scattering line. To study this,
we return to the core-only model, with velocity 3000 \kmps, and set
the rest wavelength of a single line in the core to 3000~\AA. We then
set the optical depth of the line to $\tau=1$, and proceed to exclude
this optical depth from progressively larger portions of the core's
central region. The resulting line profiles are shown in
Figure~\ref{fig:nickel_hole}. The classical rounded profile is
replaced by a flat profile, but there still exists both an emission
peak and an absorption trough.  This result corroborates the plateau
effects we studied in \S \ref{sec:one_line_outside_core} ---
Figure~\ref{fig:flux_outside_core} already showed that when a region
of large line optical depth surrounds a region of zero optical depth,
flat-topped components can appear in the spectral line profiles.

\section{Conclusions}
\label{sec:conc}
We have presented the geometric framework for a post-photospheric
model of a SN, making simple assumptions about its emissivity and
source function, in an attempt to explore the effects of
resonance-scattering in optically thick lines in SN atmospheres in
epochs where such scattering processes are rarely considered. Our
model is inspired by and similar to the Elementary Supernova model,
but the substitution of a photosphere for a transparent but
continuum-emitting core leads to significant differences in line
formation.  The most noticeable difference is that the emission peak
of a line is redshifted from its rest wavelength, even though the
model is spherically symmetric. This property may affect the
interpretation of SN spectra in post-photospheric phases, when
asymmetric effects on line formation become influential.

Our model remains mostly within the geometric confines of the
Elementary Supernova model, and choices about its properties different
than the ones made here are possible. For example, one may assume that
severe line blending in the core creates a pseudo-continuum, leading
to a source function in the shell exactly equal to that assumed in
\synow:
\[ S = W(r) J, \] where $W(r)$ is the dilution factor. 
We have also chosen to parameterize all line optical depths rather
than calculating them in any self-consistent manner, e.g., by solving
rate equations. We believe our choices in these matters have resulted
in a level of detail commensurate with the simplicity of this model.

To explore fully the other implications of resonance-scattering on
line formation at post-photospheric times in \sneia we are currently
incorporating the formalism presented in this work into SYNAPPS
\citep{synapps11}, whereby we will be able to analyze the interaction
of the receding photosphere with the emerging glowing core of \nco and
its effects on the SN spectrum, a process which has so far received
little attention.

\acknowledgements

The referee's careful eye vastly improved the content of this work,
for which we are grateful. This work was supported in part by NSF
grant AST-0707704, and US DOE Grant DE-FG02-07ER41517, and by SFB 676,
GRK 1354 from the DFG. Support for Program number HST-GO-12298.05-A
was provided by NASA through a grant from the Space Telescope Science
Institute, which is operated by the Association of Universities for
Research in Astronomy, Incorporated, under NASA contract NAS5-26555.
This research has made use of NASA's Astrophysics Data System.

\clearpage

%\bibliography{apj-jour,mystrings,refs,baron,sn1bc,sn1a,sn87a,snii,stars,rte,cosmology,gals,agn,atomdata,crossrefs,grb}

\begin{thebibliography}{48}
\expandafter\ifx\csname natexlab\endcsname\relax\def\natexlab#1{#1}\fi

\bibitem[{{Axelrod}(1980)}]{axelphd80}
{Axelrod}, T.~S. 1980, PhD thesis, California Univ., Santa Cruz.

\bibitem[{Bowers {et~al.}(1997)}]{bowersetal97}
Bowers, E. {et~al.} 1997, MNRAS, 290, 663

\bibitem[{{Branch} {et~al.}(2005){Branch}, {Baron}, {Hall}, {Melakayil}, \&
  {Parrent}}]{branchcomp105}
{Branch}, D., {Baron}, E., {Hall}, N., {Melakayil}, M., \& {Parrent}, J. 2005,
  PASP, 117, 545

\bibitem[{Branch {et~al.}(1985)Branch, Doggett, Nomoto, \&
  Thielemann}]{branch81b85}
Branch, D., Doggett, J.~B., Nomoto, K., \& Thielemann, F.-K. 1985, ApJ, 294,
  619

\bibitem[{Branch {et~al.}(1983)}]{bran81b}
Branch, D. {et~al.} 1983, ApJ, 270, 123

\bibitem[{Castor(1970)}]{castor70}
Castor, J.~I. 1970, MNRAS, 149, 111

\bibitem[{De {et~al.}(2010)De, Baron, \& Hauschildt}]{soma10a}
De, S., Baron, E., \& Hauschildt, P.~H. 2010, MNRAS, 401, 2081

\bibitem[{Dessart \& Hillier(2005)}]{dessart05a}
Dessart, L. \& Hillier, D.~J. 2005, A\&A, 437, 667

\bibitem[{{Dessart} \& {Hillier}(2010)}]{DH10a}
{Dessart}, L. \& {Hillier}, D.~J. 2010, MNRAS, 405, 2141

\bibitem[{{Dessart} \& {Hillier}(2011)}]{dh11}
---. 2011, \mnras, 410, 1739

\bibitem[{{Fesen} {et~al.}(2007){Fesen}, {H{\"o}flich}, {Hamilton}, {Hammell},
  {Gerardy}, {Khokhlov}, \& {Wheeler}}]{fesen07}
{Fesen}, R.~A., {H{\"o}flich}, P.~A., {Hamilton}, A.~J.~S., {Hammell}, M.~C.,
  {Gerardy}, C.~L., {Khokhlov}, A.~M., \& {Wheeler}, J.~C. 2007, ApJ, 658, 396

\bibitem[{Fisher {et~al.}(1997)Fisher, Branch, Nugent, \& Baron}]{fish90n97}
Fisher, A., Branch, D., Nugent, P., \& Baron, E. 1997, ApJ, 481, L89

\bibitem[{{Foley} \& {Kasen}(2011)}]{foley-kasen11}
{Foley}, R.~J. \& {Kasen}, D. 2011, ApJ, 729, 55

\bibitem[{{Foley} {et~al.}(2012)}]{foley12b}
{Foley}, R.~J. {et~al.} 2012, \apj, 752, 101

\bibitem[{{Gerardy} {et~al.}(2007)}]{gerardy07}
{Gerardy}, C.~L. {et~al.} 2007, \apj, 661, 995

\bibitem[{{Hachinger} {et~al.}(2006){Hachinger}, {Mazzali}, \&
  {Benetti}}]{hach06}
{Hachinger}, S., {Mazzali}, P.~A., \& {Benetti}, S. 2006, MNRAS, 370, 299

\bibitem[{Hatano {et~al.}(1999{\natexlab{a}})Hatano, Branch, Fisher, Baron, \&
  Filippenko}]{hatano94D99}
Hatano, K., Branch, D., Fisher, A., Baron, E., \& Filippenko, A.~V.
  1999{\natexlab{a}}, ApJ, 525, 881

\bibitem[{Hatano {et~al.}(1999{\natexlab{b}})Hatano, Branch, Fisher, Deaton, \&
  Baron}]{atlas99}
Hatano, K., Branch, D., Fisher, A., Deaton, J., \& Baron, E.
  1999{\natexlab{b}}, ApJS, 121, 233

\bibitem[{{Hillier} \& {Dessart}(2012)}]{HD12}
{Hillier}, D.~J. \& {Dessart}, L. 2012, \mnras, 424, 252

\bibitem[{{H{\"o}flich}(2003)}]{h03a}
{H{\"o}flich}, P. 2003, in Astronomical Society of the Pacific Conference
  Series, Vol. 288, Stellar Atmosphere Modeling, ed. {I.~Hubeny, D.~Mihalas, \&
  K.~Werner} (San Franciso: ASP), 185

\bibitem[{{H{\"o}flich} {et~al.}(2004){H{\"o}flich}, {Gerardy}, {Nomoto},
  {Motohara}, {Fesen}, {Maeda}, {Ohkubo}, \& {Tominaga}}]{hof03du04}
{H{\"o}flich}, P., {Gerardy}, C.~L., {Nomoto}, K., {Motohara}, K., {Fesen},
  R.~A., {Maeda}, K., {Ohkubo}, T., \& {Tominaga}, N. 2004, ApJ, 617, 1258

\bibitem[{{Jack} {et~al.}(2009){Jack}, {Hauschildt}, \& {Baron}}]{jhb09}
{Jack}, D., {Hauschildt}, P.~H., \& {Baron}, E. 2009, A\&A, 502, 1043

\bibitem[{Jeffery \& Branch(1990)}]{jb90}
Jeffery, D. \& Branch, D. 1990, in Supernovae, ed. J.~C. Wheeler \& T.~Piran
  (Singapore: World Scientific), 149

\bibitem[{Jeffery {et~al.}(1992)Jeffery, Leibundgut, Kirshner, Benetti, Branch,
  \& Sonneborn}]{jeffetal92}
Jeffery, D., Leibundgut, B., Kirshner, R.~P., Benetti, S., Branch, D., \&
  Sonneborn, G. 1992, ApJ, 397, 304

\bibitem[{{Jeffery} {et~al.}(2007){Jeffery}, {Ketchum}, {Branch}, {Baron},
  {Elmhamdi}, \& {Danziger}}]{jeffery07a}
{Jeffery}, D.~J., {Ketchum}, W., {Branch}, D., {Baron}, E., {Elmhamdi}, A., \&
  {Danziger}, I.~J. 2007, ApJS, 171, 493

\bibitem[{{Jerkstrand} {et~al.}(2011){Jerkstrand}, {Fransson}, \&
  {Kozma}}]{jerkstrand11}
{Jerkstrand}, A., {Fransson}, C., \& {Kozma}, C. 2011, \aap, 530, A45

\bibitem[{{Kasen} {et~al.}(2006){Kasen}, {Thomas}, \& {Nugent}}]{kasen06a}
{Kasen}, D., {Thomas}, R.~C., \& {Nugent}, P. 2006, ApJ, 651, 366

\bibitem[{Kozma \& Fransson(1998{\natexlab{a}})}]{kozfran98a}
Kozma, C. \& Fransson, C. 1998{\natexlab{a}}, ApJ, 496, 967

\bibitem[{Kozma \& Fransson(1998{\natexlab{b}})}]{kozfran98b}
---. 1998{\natexlab{b}}, ApJ, 497, 431

\bibitem[{{Kromer} \& {Sim}(2009)}]{kromersim09}
{Kromer}, M. \& {Sim}, S.~A. 2009, MNRAS, 398, 1809

\bibitem[{{Maeda} {et~al.}(2010){Maeda}, {Benetti}, {Stritzinger}, {R{\"o}pke},
  {Folatelli}, {Sollerman}, {Taubenberger}, {Nomoto}, {Leloudas}, {Hamuy},
  {Tanaka}, {Mazzali}, \& {Elias-Rosa}}]{maedanature10}
{Maeda}, K., {Benetti}, S., {Stritzinger}, M., {R{\"o}pke}, F.~K., {Folatelli},
  G., {Sollerman}, J., {Taubenberger}, S., {Nomoto}, K., {Leloudas}, G.,
  {Hamuy}, M., {Tanaka}, M., {Mazzali}, P.~A., \& {Elias-Rosa}, N. 2010,
  Nature, 466, 82

\bibitem[{{Maeda} {et~al.}(2011){Maeda}, {Leloudas}, {Taubenberger},
  {Stritzinger}, {Sollerman}, {Elias-Rosa}, {Benetti}, {Hamuy}, {Folatelli}, \&
  {Mazzali}}]{maeda11}
{Maeda}, K., {Leloudas}, G., {Taubenberger}, S., {Stritzinger}, M.,
  {Sollerman}, J., {Elias-Rosa}, N., {Benetti}, S., {Hamuy}, M., {Folatelli},
  G., \& {Mazzali}, P.~A. 2011, \mnras, 413, 3075

\bibitem[{{Maeda} {et~al.}(2006){Maeda}, {Nomoto}, {Mazzali}, \&
  {Deng}}]{maeda98bw06}
{Maeda}, K., {Nomoto}, K., {Mazzali}, P.~A., \& {Deng}, J. 2006, \apj, 640, 854

\bibitem[{{Maund} {et~al.}(2010){Maund}, {H{\"o}flich}, {Patat}, {Wheeler},
  {Zelaya}, {Baade}, {Wang}, {Clocchiatti}, \& {Quinn}}]{maund10a}
{Maund}, J.~R., {H{\"o}flich}, P.~A., {Patat}, F., {Wheeler}, J.~C., {Zelaya},
  P., {Baade}, D., {Wang}, L., {Clocchiatti}, A., \& {Quinn}, J. 2010, ApJ,
  725, L167

\bibitem[{{Maurer} {et~al.}(2011){Maurer}, {Jerkstrand}, {Mazzali},
  {Taubenberger}, {Hachinger}, {Kromer}, {Sim}, \& {Hillebrandt}}]{Maurer11}
{Maurer}, I., {Jerkstrand}, A., {Mazzali}, P.~A., {Taubenberger}, S.,
  {Hachinger}, S., {Kromer}, M., {Sim}, S., \& {Hillebrandt}, W. 2011, \mnras,
  418, 1517

\bibitem[{Mazzali(2001)}]{mazz90N01}
Mazzali, P. 2001, MNRAS, 321, 341

\bibitem[{Mazzali {et~al.}(1998)Mazzali, Cappellaro, Danziger, Turatto, \&
  Benetti}]{mazzali98}
Mazzali, P., Cappellaro, E., Danziger, I., Turatto, M., \& Benetti, S. 1998,
  ApJ, 499, L49

\bibitem[{Mazzali {et~al.}(1995)Mazzali, Danziger, \& Turatto}]{mazzali95}
Mazzali, P., Danziger, I.~J., \& Turatto, M. 1995, A\&A, 297, 509

\bibitem[{Mazzali {et~al.}(2005)}]{mazzali05a}
Mazzali, P. {et~al.} 2005, ApJ, 623, L37

\bibitem[{Mazzali(2000)}]{mazzcode00}
Mazzali, P.~A. 2000, A\&A, 363, 705

\bibitem[{{Mazzali} {et~al.}(1997){Mazzali}, {Chugai}, {Turatto}, {Lucy},
  {Danziger}, {Cappellaro}, {della Valle}, \& {Benetti}}]{mazz91bg97}
{Mazzali}, P.~A., {Chugai}, N., {Turatto}, M., {Lucy}, L.~B., {Danziger},
  I.~J., {Cappellaro}, E., {della Valle}, M., \& {Benetti}, S. 1997, MNRAS,
  284, 151

\bibitem[{{Mazzali} \& {Lucy}(1993)}]{ML93}
{Mazzali}, P.~A. \& {Lucy}, L.~B. 1993, \aap, 279, 447

\bibitem[{{Mazzali} {et~al.}(2011){Mazzali}, {Maurer}, {Stritzinger},
  {Taubenberger}, {Benetti}, \& {Hachinger}}]{mazz03hv11}
{Mazzali}, P.~A., {Maurer}, I., {Stritzinger}, M., {Taubenberger}, S.,
  {Benetti}, S., \& {Hachinger}, S. 2011, \mnras, 416, 881

\bibitem[{Mihalas(1978)}]{mihalas78sa}
Mihalas, D. 1978, Stellar Atmospheres (New York: W. H. Freeman)

\bibitem[{{Motohara} {et~al.}(2006){Motohara}, {Maeda}, {Gerardy}, {Nomoto},
  {Tanaka}, {Tominaga}, {Ohkubo}, {Mazzali}, {Fesen}, {H{\"o}flich}, \&
  {Wheeler}}]{moto06}
{Motohara}, K., {Maeda}, K., {Gerardy}, C.~L., {Nomoto}, K., {Tanaka}, M.,
  {Tominaga}, N., {Ohkubo}, T., {Mazzali}, P.~A., {Fesen}, R.~A.,
  {H{\"o}flich}, P., \& {Wheeler}, J.~C. 2006, \apjl, 652, L101

\bibitem[{{Pinto} \& {Eastman}(2000)}]{PE00}
{Pinto}, P.~A. \& {Eastman}, R.~G. 2000, ApJ, 530, 757

\bibitem[{{Ruiz-Lapuente} \& {Lucy}(1992)}]{rplucy92}
{Ruiz-Lapuente}, P. \& {Lucy}, L.~B. 1992, \apj, 400, 127

\bibitem[{{Thomas} {et~al.}(2011){Thomas}, {Nugent}, \& {Meza}}]{synapps11}
{Thomas}, R.~C., {Nugent}, P.~E., \& {Meza}, J.~C. 2011, \pasp, 123, 237

\end{thebibliography}

\begin{appendix}
\label{sec:appdx}

\section{$S(r)$ outside the core}
\label{apx:S_outside}

\begin{figure}
\centering \includegraphics[scale=0.45]{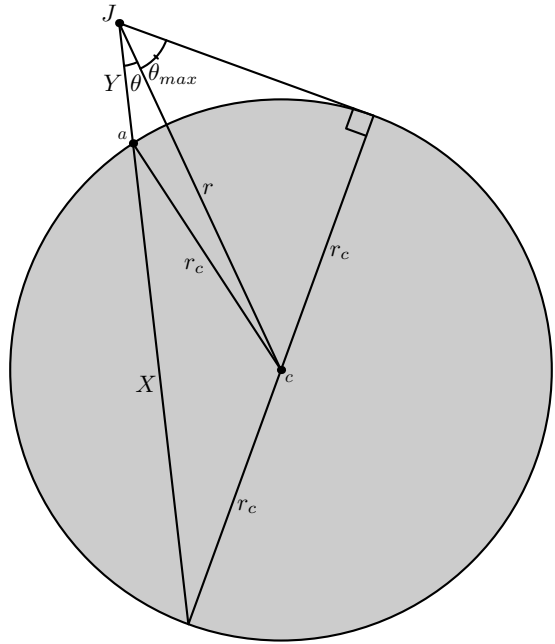}
  \caption{Geometric configuration used to calculate $S(r > \rc)$ in
    the absence of lines in the core.}
  \label{fig:SF_outside_core}
\end{figure}

In \S\ref{sec:one_line_outside_core} we calculate sample spectra for
lines forming outside the glowing core. Therefore, we calculate the
source function in the region $r > \rc$. From
Figure~\ref{fig:SF_outside_core}:
\begin{equation*}
  Y = r\mu - \left(r^2 \mu^2 + \rc^2 - r^2 \right)^{1/2}.
\end{equation*}
Then
\begin{equation}
  X = (X + Y) - Y = 2 \left(r^2 \mu^2 + \rc^2 - r^2 \right)^{1/2}.
\end{equation}
We plug the expression for $X$ into Equation~\ref{eq:jr}, noting that,
because the shell emits no continuum, the maximum value of $\theta$ is
not 1, as in the $r \leq \rc$ case, but rather
\[ \theta_{\mathrm{max}} = \cos^{-1} \left( \frac{(r^2-\rc^2)^{1/2}}{r}
\right).\] Thus,
\[ J(r) = \frac{1}{2}\int_{\mu_0}^1 2 \left( r^2 \mu^2 + \rc^2 - r^2
\right)^{1/2}\, d\mu , \] where $\mu_0$ is the argument of the inverse
cosine above.  The result is \bea J(r) &=& \frac{1}{2r}\left\{ r\rc +
(r^2-\rc^2) \ln\left[\frac{\sqrt{r^2-\rc^2}}{r+\rc}\right]
\right\}\nonumber\\ &=& \frac{1}{2r}\left\{ r\rc +
\frac{(r^2-\rc^2)}{2} \ln\left[\frac{r-\rc}{r+\rc}\right]
\right\}\nonumber.\\
  \label{eq:jrout_apx}
\eea

The mean intensities inside and outside the core are identical except
for two sign differences: one in the factor multiplying the logarithm,
and the other in the argument of the logarithm itself.  We find that
Eqs.~\ref{eq:jrin} and~\ref{eq:jrout} are continuous at the edge of
the core, as they must be, each yielding $J(r=\rc)=0.5\rc$.

\section{Two lines}
\label{apx:two_lines}
We now turn to the two-line case. We denote these lines as $R$ and
$B$, for ``red'' and ``blue.'' First, as stated previously, the source
function for the blue line, $S_B$, is given by
Equation~\ref{eq:jrin}. The source function for the red line,
$S_R(r)$, is equal to the mean intensity $J_\lambda(r)$, where
$\lambda$ is the rest wavelength of the line $\lambda_{0_R}$. The
calculation of $J$ is complicated by the fact that photons emitted by
the blue line may be scattered into resonance with the red line, while
some continuum photons from the glowing core which would, in the
absence of the blue line, redshift into resonance with the red line
are actually scattered away by the blue line. In velocity space the
region where the blue line interacts with the red line takes the form
of a ``scattering sphere'' called the \emph{common point velocity
  surface} (CPVS) \citep{jb90}. The radius of the CPVS, denoted $Y$ in
Figure~\ref{fig:full_core_with_CPS}, is given by the Doppler formula,
Equation~\ref{eq:doppler}, where one replaces $z$ in that equation
with $Y$. In \S \ref{apx:two_lines_inside_core} we study the
interaction of two lines confined to the core, and in \S
\ref{apx:two_lines_outside} we explore two lines in the shell.

\subsection{Two lines inside the core}
\label{apx:two_lines_inside_core}
\begin{figure}
\centering \includegraphics[scale=0.45]{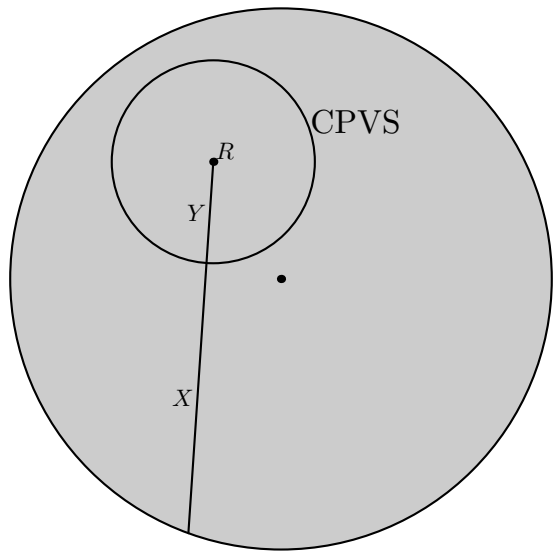}
  \caption{The different components of an intensity ray incident on
    the red line in the two-line case.}
  \label{fig:full_core_with_CPS}
\end{figure}

\begin{figure}
\centering \includegraphics[scale=0.45]{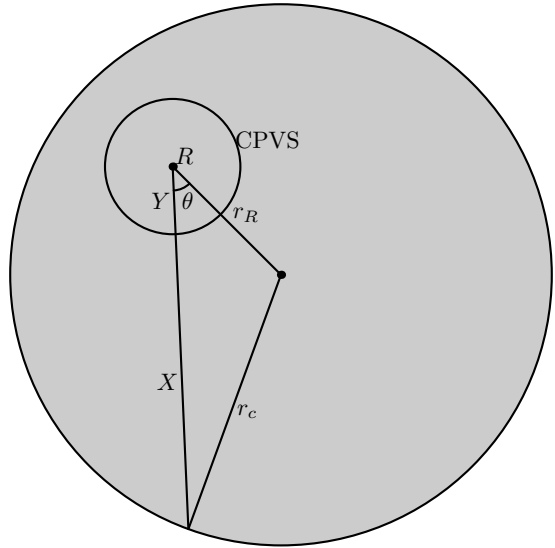}
  \caption{Explicit geometric construction of two-line configuration
    used to calculate $X~=~X(\mu)$.}
  \label{fig:calc_X_of_theta}
\end{figure}

\begin{figure}
\centering \includegraphics[scale=0.5]{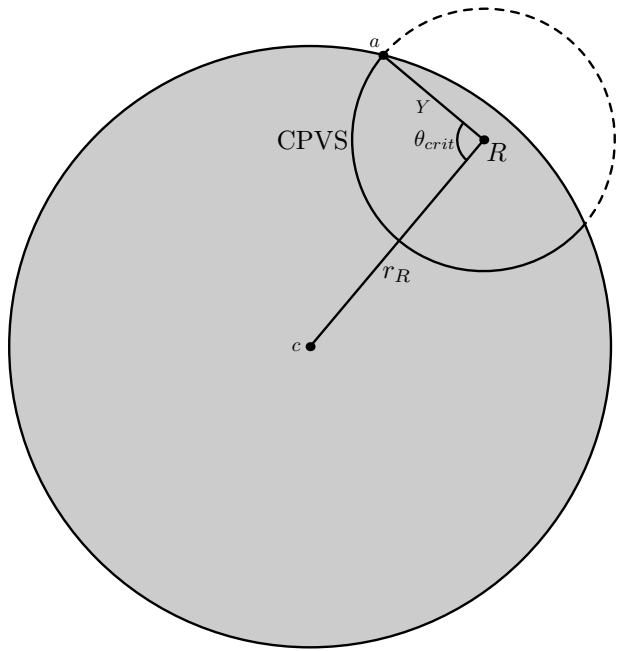}
  \caption{Geometric construction of angle
    $\theta_{\mathrm{crit}}$. The value $r_R$ is the magnitude of the
    vector $\protect\overrightarrow{cR}$, and $Y$ is that of
    $\protect\overrightarrow{aR}$.}
  \label{fig:CPS_hangs_off_core_edge_full}
\end{figure}

When both lines are confined to the core, $S_R$ contains three
components:
\begin{enumerate}
\item the intensity of continuum photons which can redshift into
  resonance with $R$, but are scattered away when they reach the CPVS;
\item photons emitted by $B$ along the CPVS which redshift into
  resonance with $R$;
\item continuum photons which form \emph{inside} the CPVS and
  therefore redshift into resonance with $R$ without interacting with
  $B$.
\end{enumerate}
Segments 1 and 3 are labeled $X$ and $Y$ respectively in
Figure~\ref{fig:full_core_with_CPS}.
Mathematically we write this as
\begin{equation}
  \label{eq:S_R}
  S_R = \frac{1}{2} \int_{-1}^1 X e^{-\tau_B} d\mu + \frac{1}{2}
  \int_{-1}^1 S_B (1-e^{-\tau_B}) d\mu + \frac{1}{2} \int_{-1}^1 Y
  d\mu.
\end{equation}
To calculate $X = X(\mu)$ we refer to
Figure~\ref{fig:calc_X_of_theta}.
Defining the triangle $ABC$ in Figure~\ref{fig:calc_X_of_theta}, we
can use the law of cosines to find:
\begin{equation}
\rc^2 = (X+Y)^2 +r_R^2 - 2 (X+Y)r_R\mu
\end{equation}
where we have used the fact that $X$ and $Y$ are co-linear and $\mu =
\cos\theta$. Solving for $X$,
\begin{equation}
X = (r_R\mu - Y) \pm \sqrt{(r_R\mu - Y)^2 + \rc^2 - r_R^2 + 2Y
  r_R\mu}.
\label{eqn:xofmu1}
\end{equation}
Expanding the square root yields
\begin{equation}
\sqrt{(r_R\mu - Y)^2 + \rc^2 - r_R^2 + 2Y r_R\mu} = \sqrt{\rc^2 -
  r_R^2(1-\mu^2) + Y^2}.
\end{equation}
When $-1 \leq \mu \leq Y/r_R$, we see that $(r_R\mu - Y)< 0$ and we
must take the positive root in Equation~\ref{eqn:xofmu1}. However,
when $Y/r_R < \mu \leq 1$ we see from Equation~\ref{eqn:xofmu1} that
$\mu > 0$ and thus the term in the square root is larger than the
expression $r_R\mu - Y$ and so the positive root is also
correct. Hence,
\begin{equation}
X = (r_R\mu - Y) + \sqrt{\rc^2 - r_R^2(1-\mu^2) + Y^2},
\end{equation} for $-1 \leq \mu \leq 1$.
Before proceeding we note that if the CPVS extends past the edge of
the core, that is, if $r_R + Y > \rc$, then there exists a critical
angle $\theta_{\mathrm{crit}}$ for which $X$ becomes undefined if
$\theta > \theta_{\mathrm{crit}}$. Its value is
\begin{equation}
  \label{eq:mu_crit}
  \mu_{\mathrm{crit}} = \frac{Y^2 + r_R^2 - 1}{2 Y r_R},
\end{equation}
where $\mu_{\mathrm{crit}} \equiv \cos \theta_{\mathrm{crit}}$.  When
integrating to find the contribution of the CPVS to the source
function of the red line, $S_R$, we must stop the integration at this
angle.  This limit is depicted in
Figure~\ref{fig:CPS_hangs_off_core_edge_full}.

\begin{figure}
\centering \includegraphics[scale=0.55]{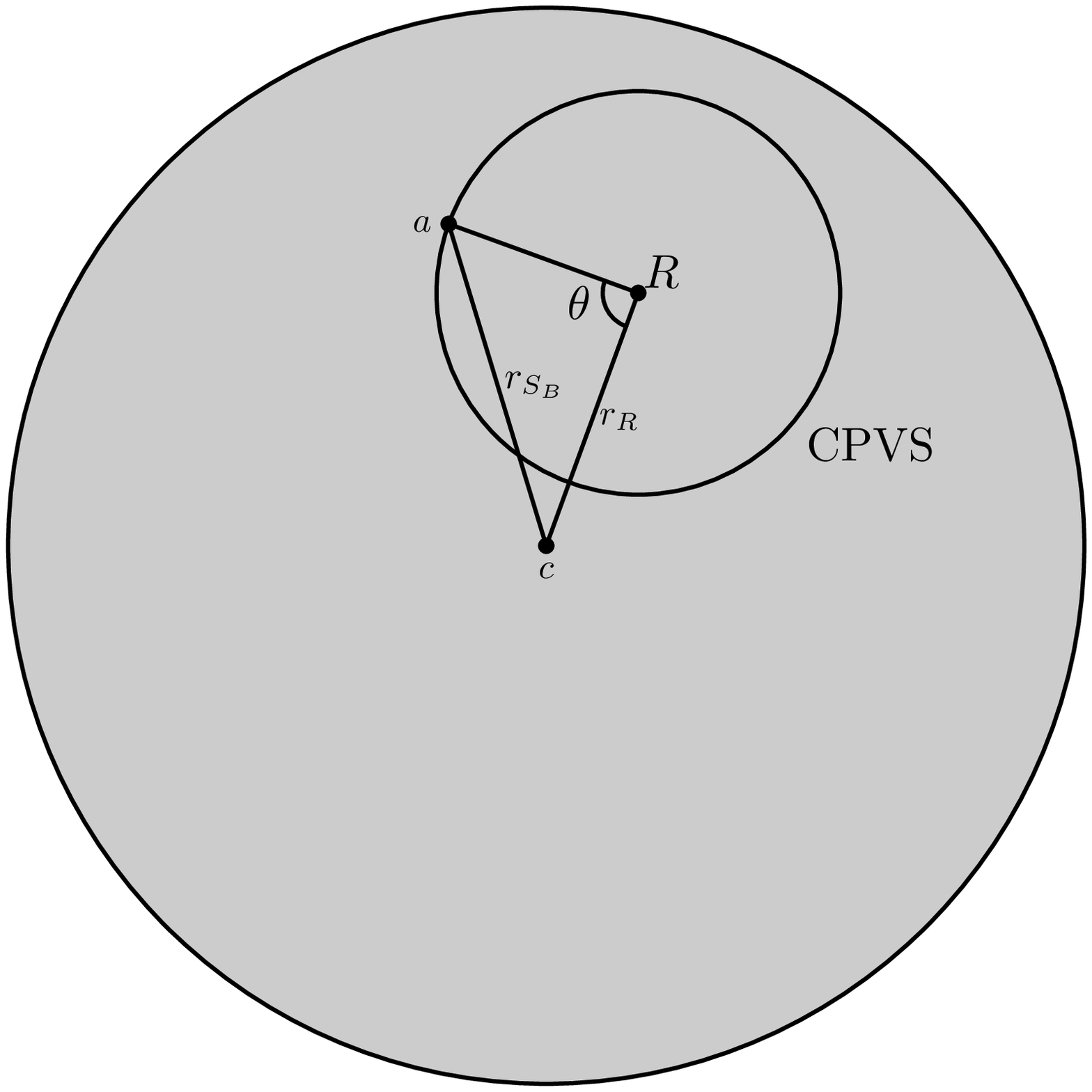}
  \caption{Calculation of location of CPVS, labeled $r_{S_B}$, with
    respect to the center of the SN. The quantity $r_R$ is the
    magnitude of the vector $\protect\overrightarrow{cR}$, and
    $r_{S_B}$ is that of $\protect\overrightarrow{ac}$.}
  \label{fig:r_S_B}
\end{figure}

\begin{figure}
\centering \includegraphics[scale=0.65]{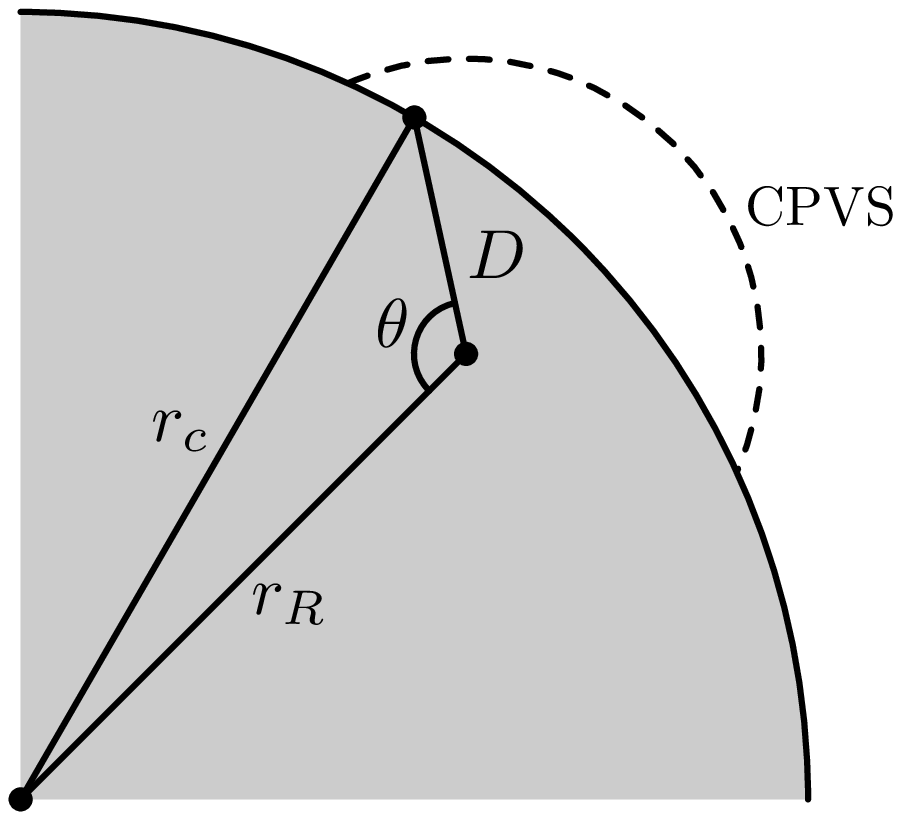}
  \caption{Geometric configuration of truncated intensity ray $D = D(\mu)$.}
  \label{fig:CPS_hangs_off_core_edge_part}
\end{figure}

Next, we must calculate $r_{S_B} = r_{S_B}(\mu)$, the location of the
CPVS, as shown in Figure~\ref{fig:r_S_B}.
Again through vector addition arguments we find
\begin{equation}
   r_{S_B}(\mu) = (r_R^2 + Y^2 - 2 r_R Y \mu)^{1/2}.
\end{equation}
It is at this location $r_{S_B}$ that both $\tau_B$ and $S_B$ in
Equation~\ref{eq:S_R} are evaluated. Lastly we turn to the calculation
of $Y$ in Figure~\ref{fig:full_core_with_CPS}. If the entire CPVS fits
inside the core, we may use the Doppler formula,
Equation~\ref{eq:doppler}, to calculate $Y$ for all $\theta$. However,
if part of the CPVS extends past the edge of the core then $Y$ takes a
slightly more complicated form. This latter case is shown in
Figure~\ref{fig:CPS_hangs_off_core_edge_part}, where $\theta_1 >
\theta_{\mathrm{crit}}$.
We calculated this angle in Equation~\ref{eq:mu_crit}.  If $\theta >
\theta_{\mathrm{crit}}$ then, from
Figure~\ref{fig:CPS_hangs_off_core_edge_part},
\begin{equation}
   D(\mu) = r_R \mu \pm (r_R^2 \mu^2 + \rc^2 - r_R^2)^{1/2}.
\end{equation}
Since $D$ is a length and must always satisfy $D \geq 0$ we can rule
out immediately the ``minus'' solution, since it is negative for all
possible values of $(r_R, Y, \mu)$. Thus we take the positive root:
\begin{equation}
  \label{eq:D_of_mu}
  D(\mu) = r_R \mu + (r_R^2 \mu^2 + \rc^2 - r_R^2)^{1/2} \quad \quad
  \quad -1 \leq \mu \leq \mu_{\mathrm{crit}}.
\end{equation}

\subsection{Two lines outside the core}
\label{apx:two_lines_outside}
\begin{figure}
  \includegraphics[scale=1.0]{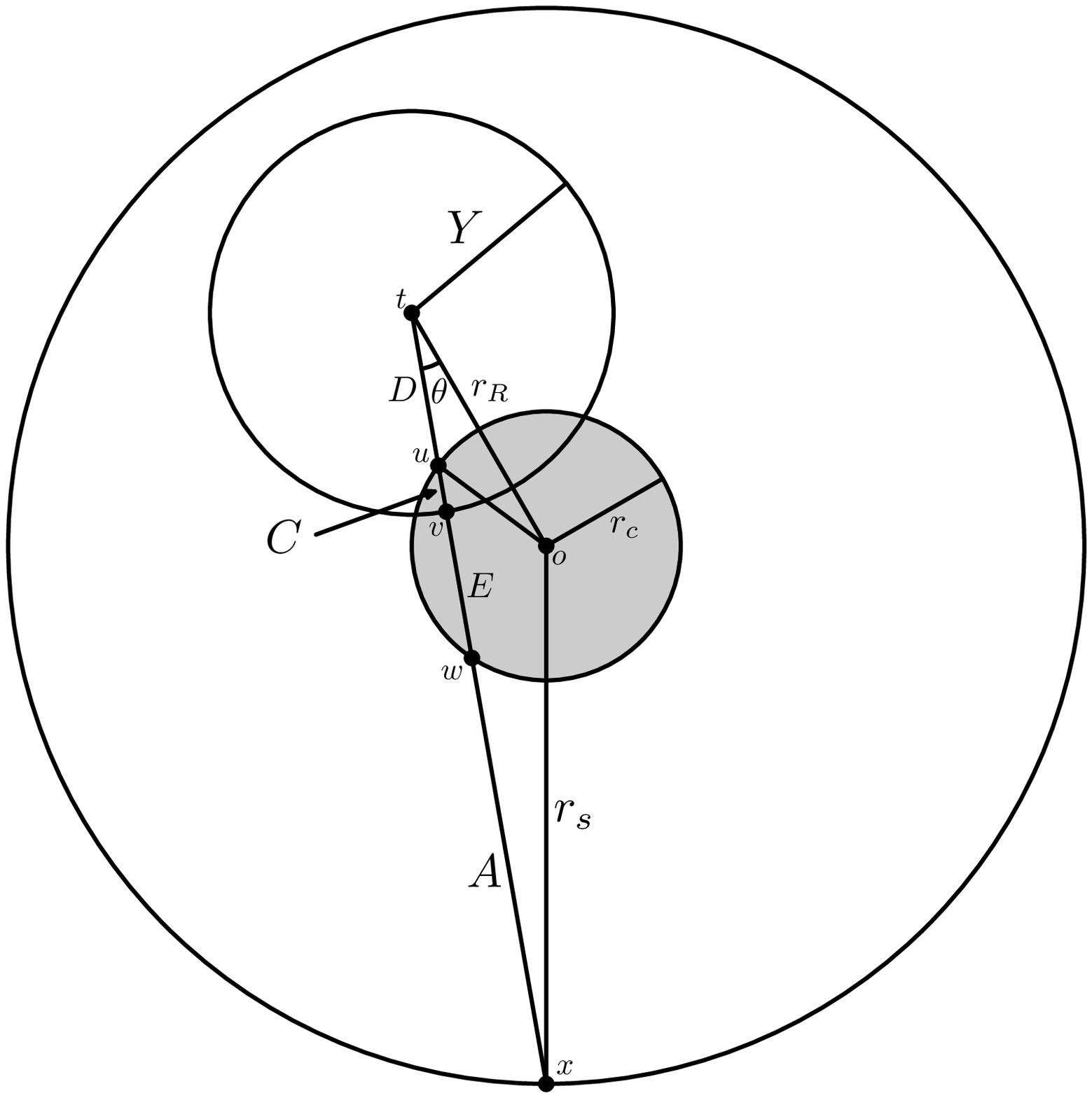}
  \caption{Case I geometry. The value of $A$ corresponds to the
    magnitude of the vector $\protect\overrightarrow{xw}$, $E$ to
    $\protect\overrightarrow{wv}$, $C$ to
    $\protect\overrightarrow{vu}$, $D$ to
    $\protect\overrightarrow{ut}$, $\rc$ to
    $\protect\overrightarrow{ou}$, and $r_s$ to
    $\protect\overrightarrow{xo}$.  Only $E$ and $C$ contribute to the
    intensity of the ray since $A$ and $D$ lie outside the emitting
    core.}
  \label{fig:case1} 
\end{figure}

\begin{figure}
  \includegraphics[scale=1.0]{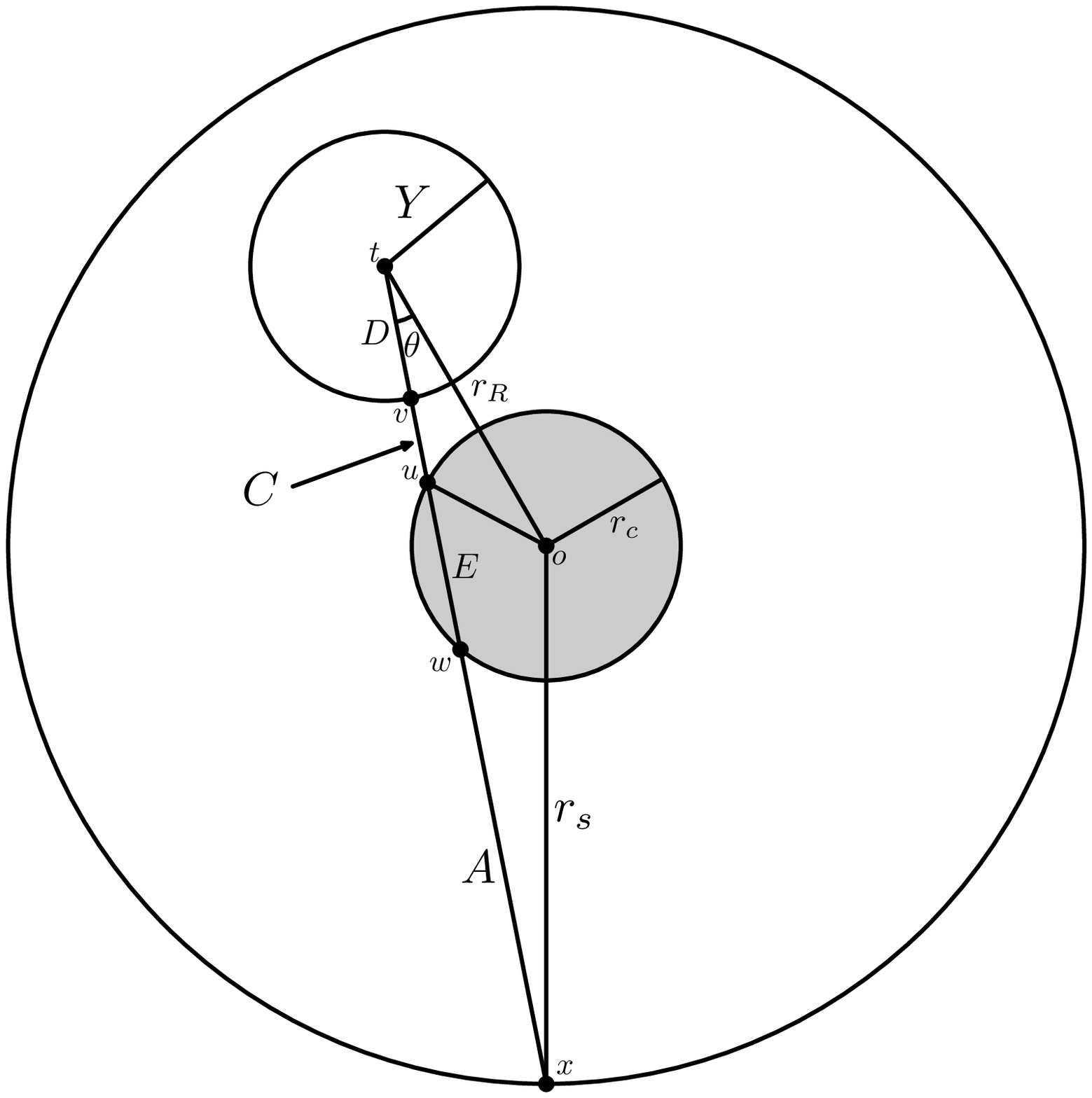}
  \caption{Case II geometry. The value of $A$ corresponds to the
    magnitude of the vector $\protect\overrightarrow{xw}$, $E$ to
    $\protect\overrightarrow{wu}$, $C$ to
    $\protect\overrightarrow{uv}$, $D$ to
    $\protect\overrightarrow{ut}$, $\rc$ to
    $\protect\overrightarrow{ou}$, and $r_s$ to
    $\protect\overrightarrow{xo}$.  Only $E$ contributes to the
    intensity of the ray since $A$, $C$, and $D$ lie outside the
    emitting core.}
  \label{fig:case2} 
\end{figure}

\begin{figure}
  \includegraphics[scale=1.0]{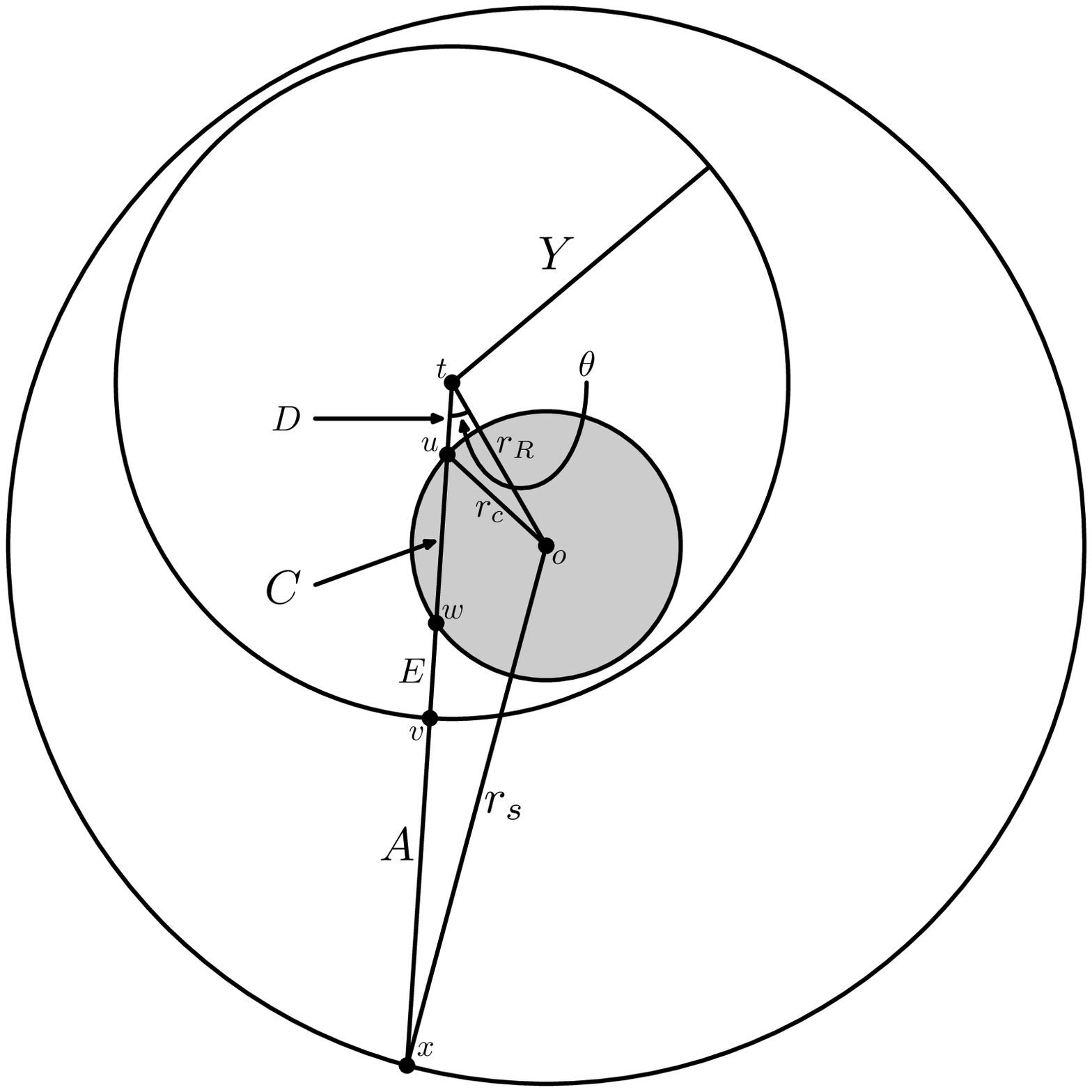}
  \caption{Case III geometry. The value of $A$ corresponds to the
    magnitude of the vector $\protect\overrightarrow{xv}$, $E$ to
    $\protect\overrightarrow{vw}$, $C$ to
    $\protect\overrightarrow{wu}$, $D$ to
    $\protect\overrightarrow{ut}$, $\rc$ to
    $\protect\overrightarrow{ou}$, and $r_s$ to
    $\protect\overrightarrow{xo}$.  Only $C$ contributes to the
    intensity of the ray since $A$, $E$, and $D$ lie outside the
    emitting core.}
  \label{fig:case3}
\end{figure}

In the case that we have a resonance line that is strong under cold
conditions, for example, certain lines of Ca II or Mg II, we may want
to consider multiple lines forming outside the core.
Figure~\ref{fig:case1} illustrates the most complex case. If $\theta >
\theta_{\mathrm{crit}}$, where $\mu_{\mathrm{crit}} =
\cos(\theta_{\mathrm{crit}}) = 1 - (\rc/r_R)^2$, then the
characteristic in that direction does not intersect the core and
accumulates no intensity, but for the case that $ \mu \geq
\mu_{\mathrm{crit}}$ then the characteristic is given as shown in
Cases I--III, (Figures~\ref{fig:case1}--\ref{fig:case3}).

For Case I, Figure~\ref{fig:case1}, we find
\begin{equation}
D=r_R\cos\theta-\sqrt{\rc^2-r_R^2\sin^2\theta},
\end{equation}
\begin{equation}
C= Y-D = Y-r_R\cos\theta+\sqrt{\rc^2-r_R^2\sin^2\theta},
\end{equation}
\begin{equation}
E=2(r_R\cos\theta-D)-C =
\sqrt{\rc^2-r_R^2\sin^2\theta}-Y+r_R\cos\theta,
\end{equation}
and
\begin{equation}
A=\sqrt{r_s^2-r_R^2\sin^2\theta}-\sqrt{\rc^2-r_R^2\sin^2\theta}.
\end{equation}
For Case II, Figure~\ref{fig:case2}, we have
\begin{equation}
E=2\sqrt{\rc^2-r_R^2\sin^2\theta}
\end{equation}
\begin{equation}
C=r_R\cos\theta -Y -\sqrt{\rc^2-r_R^2\sin^2\theta}
\end{equation}
\begin{equation}
A=\sqrt{r_s^2-r_R^2\sin^2\theta}-\sqrt{\rc^2-r_R^2\sin^2\theta}
\end{equation}
and
\begin{equation}
D=Y.
\end{equation}
For Case III, Figure~\ref{fig:case3},
\begin{equation}
D= r_R\cos\theta -\sqrt{\rc^2-r_R^2\sin^2\theta}
\end{equation}
\begin{equation}
C= 2\sqrt{\rc^2-r_R^2\sin^2\theta}
\end{equation}
\begin{equation}
E=Y-D-C= Y-\sqrt{\rc^2-r_R^2\sin^2\theta}-r_R\cos\theta.
\end{equation}
 and
\begin{equation}
A=\sqrt{r_s^2-r_R^2\sin^2\theta}+r_R\cos\theta-Y.
\end{equation}

Then $J$ for Case I is:
\begin{equation}
J = \frac{1}{2}\left\{ \int_{-1}^{\mu_{\mathrm{crit}}}
S_B(1-e^{-\tau_B}) \,d\mu + \int_{\mu_{\mathrm{crit}}}^1 \left[
  Ee^{-\tau_B} + S_B(1-e^{-\tau_B}) + C \right] \,d\mu \right\}.
\end{equation}
For Case II, $J$ is:
\begin{equation}
J = \frac{1}{2}\left\{ \int_{-1}^{\mu_{\mathrm{crit}}}
S_B(1-e^{-\tau_B}) \,d\mu + \int_{\mu_{\mathrm{crit}}}^1 \left[
  Ee^{-\tau_B} + S_B(1-e^{-\tau_B}) \right] \,d\mu \right\}.
\end{equation}
Finally, for Case III,
\begin{equation}
J = \frac{1}{2}\left\{ \int_{-1}^{\mu_{\mathrm{crit}}}
S_B(1-e^{-\tau_B}) \,d\mu + \int_{\mu_{\mathrm{crit}}}^1 \left[ E +
  S_B(1-e^{-\tau_B}) \right] \,d\mu \right\}
\end{equation}
  
The extension of the two-line case to the $N$-line case is
straightforward and is implemented most easily in a recursive fashion,
moving along wavelength as the value of $r_R$ increases.
  
\subsection{Calculation of emergent intensity}
Given that the flux integral in Equation~\ref{eq:flux} can be written
as an integral over impact parameter $p$, we concern ourselves here
with the calculation of the emergent intensity of rays with constant
$p$, denoted $I_\lambda(p)$. We first calculate the domain
$[z_{\mathrm{min}}, z_{\mathrm{max}}]$ over which a ray with given $p$
exists:
\begin{eqnarray*}
  z_{\mathrm{min}} &=& -(\rc^2 - p^2)^{1/2} \\ z_{\mathrm{max}} &=&
  +(\rc^2 - p^2)^{1/2}
\end{eqnarray*}
We then use the Doppler formula to establish the wavelength domain
$[\lambda_{\mathrm{min}}, \lambda_{\mathrm{max}}]$ over which it is
possible for a particular wavelength point $\lambda$ to be in
resonance in the core, given $p$: \bea \lambda_{\mathrm{min}} &=&
\frac{\lambda}{1 + z_{\mathrm{max}}
  \left(\frac{v_{\mathrm{core}}}{c}\right)} \\ \lambda_{\mathrm{max}}
&=& \frac{\lambda}{1 + z_{\mathrm{min}}
  \left(\frac{v_{\mathrm{core}}}{c}\right)}.  \eea All lines with rest
wavelength $\lambda_0$ which lie in this domain will be in resonance
with the ray $I_\lambda(p)$ at some location in the core.

Now consider a ray with two lines which both fall in the domain
$[\lambda_{\mathrm{min}}, \lambda_{\mathrm{max}}]$. The ray redshifts
into resonance with the blue line ``before'' (at larger $z$, closer to
the back of the core) redshifting into resonance with the red line; we
denote these locations $z_B$ and $z_R$, respectively, where $z_B >
z_R$. The segment of the ray between the back edge of the core and
$z_B$ has intensity $(\rc^2-p^2)^{1/2}-z_B$. However it will redshift
into resonance with \emph{both} lines before emerging from the core,
so it acquires two attenuation terms, $e^{-\tau_B}$ and $e^{-\tau_R}$,
where $\tau_B$ and $\tau_R$ are evaluated at $z_B$ and $z_R$,
respectively. The continuum segment between $z_R$ and $z_B$ has length
$z_B - z_R$ and is attenuated only by the red line. The front-most
continuum piece, between $z_R$ and the front of the core, is
unaffected by scattering and has intensity $(\rc^2-p^2)^{1/2}+z_R$.

The source functions of the red and blue lines also contribute to the
total emergent intensity. The contribution from each is
$S_i(1-e^{-\tau_i})$, where $i \in \{R, B\}$. However the blue line's
contribution will be attenuated when it redshifts into resonance with
the red line, and so it is receives the usual $e^{-\tau_R}$
multiplicative factor.  Therefore the total emergent intensity is \bea
I_\lambda(p) &=& ((\rc^2 - p^2)^{1/2} - z_B) e^{-\tau_B} e^{-\tau_R} +
S_B(1-e^{-\tau_B}) e^{-\tau_R} + (z_B - z_R) e^{-\tau_R} \nonumber
\\ && + S_R(1-e^{-\tau_R}) + (\rc^2-p^2)^{1/2}+z_R.  \eea If for a
particular ray one of the two lines is outside
$[\lambda_{\mathrm{min}}, \lambda_{\mathrm{max}}]$, the result for
$I_\lambda(p)$ reduces to the one-line form for $I_\lambda(p)$ of
Equation~\ref{eq:coreline} or Eqs.~\ref{eqn:I_out1}, \ref{eqn:I_out2},
and \ref{eqn:I_out3} for lines outside the core.  It is important to
understand that if, for some impact parameter $p$, the blue line has
wavelength outside $[\lambda_{\mathrm{min}}, \lambda_{\mathrm{max}}]$,
such that $I_\lambda (p)$ is given by Equation~\ref{eq:coreline}, the
source function $S = S_R$ evaluated in that equation may nevertheless
contain scattering effects of the blue line, as long as at least some
portion of the CPVS for the blue line lines within the core.

If both lines fall outside this domain the emergent intensity becomes
the pure continuum result.

\end{appendix}

\end{document}